\documentclass[11pt,twoside]{article}

\usepackage{epsfig,amsfonts,color}
\usepackage{amsmath}

\usepackage{amssymb, palatino, geometry,url}
\usepackage{algorithmic}
\usepackage[noresetcount,lined,boxed]{algorithm2e} 

\usepackage[colorlinks=true,linkcolor=blue,citecolor=blue,urlcolor=blue]{hyperref}

\usepackage{fancyhdr}
\usepackage{amsthm}
\newtheorem{theorem}{Theorem}

\usepackage{appendix}
\usepackage{multirow}

\usepackage{caption}
\usepackage{subcaption}

\begin{document}

\title{Pricing and Remunerating Electricity Storage Flexibility \\ Using Virtual Links}

\author{Weiqi Zhang${}^{\ddag}$, Philip A. Tominac${}^{\ddag}$, and Victor M. Zavala${}^{\ddag}$\thanks{Corresponding Author: victor.zavala@wisc.edu}\\
  {\small ${}^{\ddag}$Department of Chemical and Biological Engineering}\\
 {\small \;University of Wisconsin-Madison, 1415 Engineering Dr, Madison, WI 53706, USA}}
 \date{}
\maketitle

\begin{abstract}
Ambitious renewable portfolio standards motivate the incorporation of energy storage resources (ESR) as sources of flexibility. While the United States government aims to promote ESR participation in electricity markets, work on market designs for properly remunerating ESRs is still lacking. In this paper, we propose a new energy market clearing framework that incorporates ESR systems. The new market design is computationally attractive in that it avoids mixed-integer formulations and formulations with complementarity constraints. Moreover, compared to previous market designs, our market decomposes the operations of ESRs using the concept of virtual links, which capture the transfer of energy across time. The virtual link representation reveals economic incentives available for ESR operations and sheds light into how electricity markets should remunerate ESRs. We also explore the role of ESR physical parameters on market behavior; we show that, while energy and power capacity defines the amount of flexibility each ESR can provide, storage charge/discharge efficiencies play a fundamental role in ESR remuneration and in mitigating market price volatility. We use our proposed framework to analyze the interplay between ESRs and independent system operators (ISOs) and to provide insights into optimal deployment strategies of ESRs in power grids. 
\end{abstract}


\section{Introduction}

The power grid in the United States is undergoing significant changes driven by increasing adoption of renewable energy. Multiple states, such as California, Colorado, and Virginia, have set an ambitious renewable portfolio standards of 100\% by 2050 or earlier \cite{rps}. It has been shown that electricity storage resources (ESR) offer significant flexibility potential \cite{sioshansi2009estimating, lund2020improving}; however, current electricity markets (ISOs) are not explicitly designed to enable participation of ESRs. In 2018, the Federal Energy Regulatory Commission (FERC) released Order 841 that aimed to remove barriers to wholesale electricity market participation of ESR systems \cite{ferc2018}. A detailed analysis of FERC Order 841 has been done recently \cite{smith2019enabling}. Various participation models and market rules for ESRs have been proposed for different markets (energy, ancillary and capacity) since the release of the FERC order \cite{chandra2020ferc}. 
\\

Several research works have attempted to design new energy market clearing models to incorporate ESR participants. This is important especially for large-scale ESR systems in order to understand potential for operational feasibility and to understand potential effects of ESR systems on electricity market prices. In terms of coordinated (centralized) market design (where ESR systems act as price-takers), work in \cite{parvar2019analysis} presents a basic modeling framework for electricity markets with ESR systems that forms the basis for CAISO electricity market regulations. The work in \cite{chen2020improved}
proposes a market clearing formulation that captures ESR systems with flexible terminal charge conditions that enables higher flexibility potential. The concept of financial storage rights (FSR) under coordinated market setting is proposed in various previous works, based on the dual prices of ESR physical constraints \cite{taylor2014financial} and locational marginal prices \cite{munoz2017financial}. The concept of physical storage rights (PSRs) is proposed as an alternative market product for ESR systems, where instead of being managed by the market operator, ESRs sell their energy and power capacities and dispatch actions to other market participants \cite{he2011novel, brijs2016auction, thomas2020local}. The incorporation of ESR systems in peer-to-peer markets in a smart grid is studied in \cite{luth2018local}.
\\

In addition, much research work on market design for ESR explores price-making behavior of these systems using bi-level programming approaches. For instance, a bi-level program between ISO and storage is formulated to study the optimal strategy for ESR participation, where the lower level proposes a market clearing formulation treating ESR as load and supply \cite{pandvzic2015energy}. That work shows that ESR systems prefer locations and bidding strategies that do not incur large changes in market prices (so they can act as price-takers). Work of \cite{hartwig2015impact} applies a similar bi-level framework to study the effect of ownership arrangements of ESR. A bi-level programming framework is proposed in \cite{mohsenian2015coordinated} for large-scale ESRs that are geo-distributed and coordinated. Work of \cite{huang2017market} provides a comparative analysis on market mechanisms for ESR systems with different level of coordination. 
\\

In this work, we propose a new energy market design that aims to properly incentivize flexibility provided by ESR systems. The new market design applies a robust bound on ESR operations so that complementarity constraints are not necessary. We discuss conditions where complementarity is violated in traditional market formulations for ESRs, and demonstrate that the new market formulation ensures that complementarity is satisfied at the expense of reducing operational space for ESRs. In addition, the new market design decomposes charging/discharging operations of ESR systems into temporal energy transfer and net-charging/discharging. Charging operation is treated as a pure service that should be remunerated, instead of utility gained by the ESR (as for load modeling). Energy transfer is captured using the notion of virtual links (transfer/shifting of energy across time). This reveals how our market framework incentivizes flexible operations of ESR via temporal price differences (volatility) and reveals mechanism for remunerating ESRs. We also explore the effect of different physical parameters on flexibility provision of each ESR. Specifically, we show that the energy and power capacity determine the amount of flexibility each ESR can provide. However, the efficiency of ESR determines the amount of reduction in price volatility. Via numerical experiments, we explore optimal investment and operational strategies for ESRs and their interplay with ISO goals. 
\\

The paper is structured as follows. Section \ref{sec:storage_model} details the operation model of ESRs we use in our market clearing model. Section \ref{sec:markets} details our market designs starting from the basic market clearing model. Section \ref{sec:cases} demonstrate our market clearing model using numerical experiments. Section \ref{sec:conclusion} concludes the paper. 

\section{ESR Operation Model}
\label{sec:storage_model}

\begin{figure}[h!]
    \centering
    \includegraphics[width=0.4\textwidth]{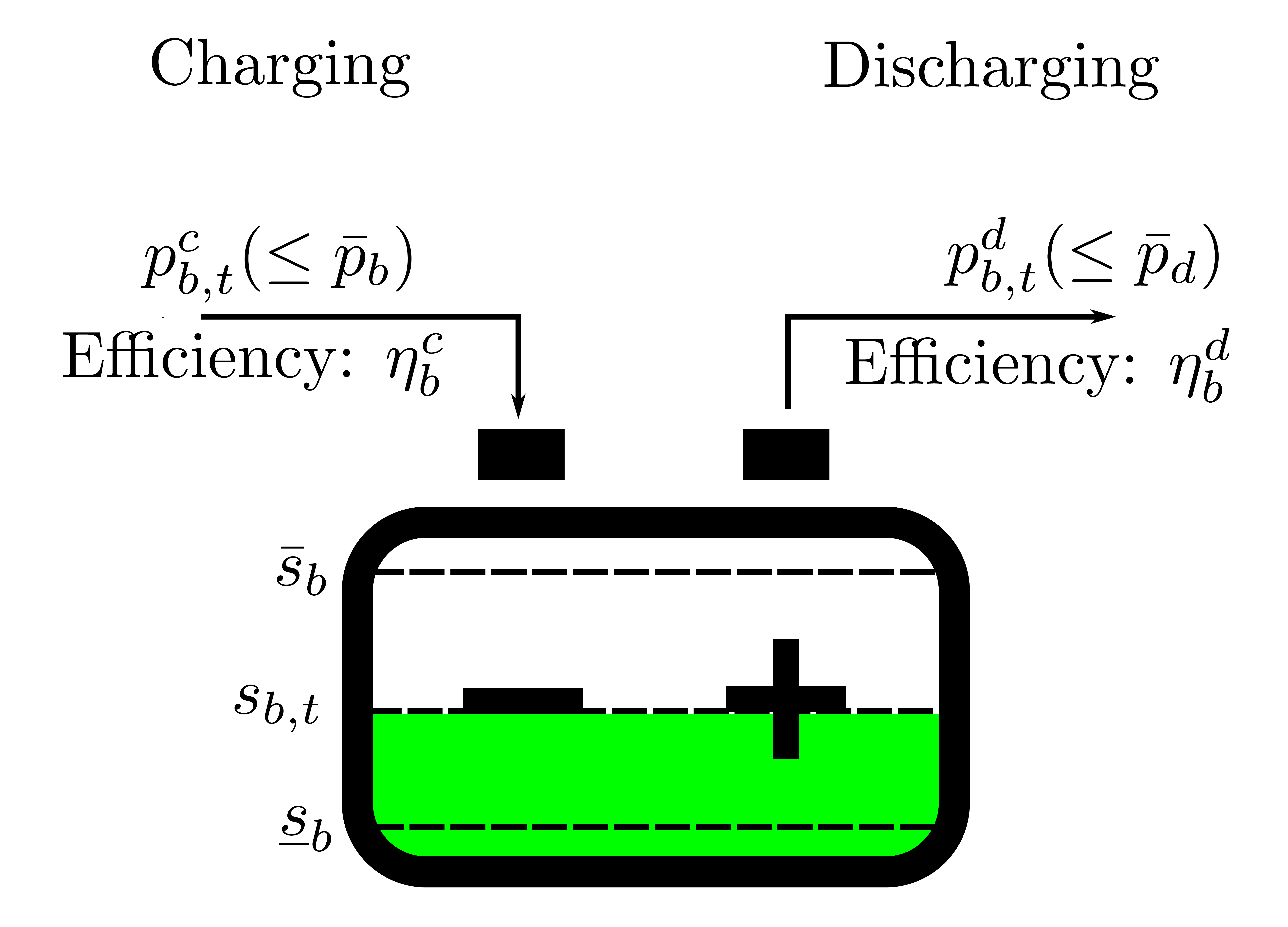}
    \caption{\small Schematics of the ESR operation model at time $t$.}
    \label{fig:storage_model}
\end{figure}

In this section we provide a description of the operational model for the ESR systems that we consider.  The operation model is based on the work of \cite{parvar2019analysis} and attempts to capture key constraints commonly seen in previous works on modeling and market incorporation of ESR systems. Notations of the operational model is summarized in Figure \ref{fig:storage_model}.

The storage operational model keeps track of the states and operations of an ESR system $b$ over a discretized time horizon $\mathcal{T} := \{1,2,...,T\}$ to ensure they obey certain physical constraints. Specifically, we let $s_{b,t}\in \mathbb{R}$ be the state-of-charge at the end of time interval $t$, $p^c_{b,t} \in \mathbb{R}_+$ be the charging power of $b$ during time interval $t$, and $p^d_{b,t} \in \mathbb{R}_+$ be the discharging power of $b$ during time interval $t$. The ESR system $b$ is defined by a set of physical parameters: 
\begin{itemize}
    \item $\eta^c_b \in [0,1]$: charging efficiency
    \item $\eta^d_b \in [0,1]$: discharging efficiency
    \item $\underline{s}_b \in \mathbb{R}_+$: minimum state of charge
    \item $\bar{s}_b \in \mathbb{R}_+$: maximum state of charge
    \item $s_{b,0} \in [\underline{s}_b,\bar{s}_b]$: state of charge at time 0
    \item $\bar{p}_b \in \mathbb{R}_+$: power capacity for charging/discharging
\end{itemize}
The storage model is formulated as follows:
\begin{subequations}
\label{eq:storage}
\begin{gather}
    s_{b,t} = s_{b,t-1} + \eta^c_b p^c_{b,t} - \frac{1}{\eta^d_b} p^d_{b,t}, \,\, t \in \mathcal{T} \label{eq:storage_balance} \\
    \underline{s}_b \leq s_{b,t} \leq \bar{s}_b, \,\, t \in \mathcal{T} \label{eq:storage_s_range}\\
    p^c_{b,t} + p^d_{b,t} \leq \bar{p}_{b}, \,\, t \in \mathcal{T} \label{eq:storage_power_cap} \\
    0 \leq p^c_{b,t} \perp p^d_{b,t} \geq 0, \,\, t \in \mathcal{T} \label{eq:storage_perp} \\
    s_{b,T} \geq s_{b,0} \label{eq:storage_final}
\end{gather}
\end{subequations}
where $\eta^c_b, \eta^d_b \in [0,1], \underline{s}_b,\bar{s}_b, p_{b} \in \mathbb{R}_+$ are physical parameters of the storage system $b$. Equation \eqref{eq:storage_balance} captures dynamics of state of charge over time. We implicitly assume that the charging and discharging time is 1 hour (so $\Delta t$ is not included in the model). Constraint \eqref{eq:storage_s_range} captures the state of charge capacity constraints. Constraint \eqref{eq:storage_power_cap} captures power capacity constraints. Constraint \eqref{eq:storage_perp} is the complementarity constraint for charging and discharging decisions, i.e. during each time interval an ESR must be either charging or discharging, but not both. Constraint \eqref{eq:storage_final} captures the terminal condition for state of charge, which is not physically necessary but rather a market operation requirement. One can capture the charging/discharging complementarity by using binary variables, which will make the model mixed-integer in nature. Regardless of the modeling approach used, the presence of complementarity makes the storage operation model nonconvex and combinatorial in nature, which makes incorporating storage systems in market clearing models challenging.  
\\

Equation \eqref{eq:storage_balance} can be aggregated to the following equivalent equation:
\begin{equation}
    s_{b,t} = s_{b,0} + \eta^c_b \sum_{t' = 1}^t p^c_{b,t'} - \frac{1}{\eta^d_b} \sum_{t' = 1}^t p^d_{b,t'}
\end{equation}
This allows us to write the storage model \eqref{eq:storage} purely in terms of $p^c$ and $p^d$:
\begin{subequations}
\label{eq:storage_nos}
\begin{gather}
    \Delta \underline{s}_{b,t} \leq \eta^c_b \sum_{t' = 1}^t p^c_{b,t'} - \frac{1}{\eta^d_b} \sum_{t' = 1}^t p^d_{b,t'} \leq \Delta \bar{s}_b, \,\, t \in \mathcal{T} \label{eq:storage_nos_energy_cap}\\
    p^c_{b,t} + p^d_{b,t} \leq \bar{p}_{b}, \,\, t \in \mathcal{T} \label{eq:storage_nos_power_cap} \\
    0 \leq p^c_{b,t} \perp p^d_{b,t} \geq 0, \,\, t \in \mathcal{T} \label{eq:storage_nos_perp}
\end{gather}
\end{subequations}
where $\Delta \underline{s}_{b,t} := \underline{s}_b - s_{b,0}$ for $t < T$, $\Delta \underline{s}_{b,T} := 0$, and $\Delta \bar{s}_b := \bar{s}_b - s_{b,0}$. Inequality \eqref{eq:storage_final} is merged into the left-hand side of \eqref{eq:storage_nos_energy_cap} via the definition of $\Delta \underline{s}_{b,T}$.

\section{Market Formulation with Storage}
\label{sec:markets}
The market clearing formulations that we propose apply the concept of virtual links to capture ESR systems. Virtual links were first proposed to capture load shifting flexibility from data centers \cite{zhang2020flexibility, zhang2021electricity} and are not standard in the power systems literature; as such, we introduce a family of formulations with increasing complexity. 

\subsection{Notation and Terminology}

We begin our discussion by introducing basic notation for a space-time network without considering ESR systems. The detailed setup and derivation for the space-time market clearing formulation is discussed in \cite{zhang2021electricity}. Here we give a brief review of the setup. The market considers, over time horizon $\mathcal{T} := \{1,2,...,T\}$, a set of suppliers (owners of power plants) $\mathcal{S}$ and consumers (owners of loads) $\mathcal{D}$. Each participant is connected to a transmission network comprised of geographical nodes $\mathcal{N}$ and transmission lines $\mathcal{L}$ (owned by transmission service providers).
\\

Each supplier $i \in \mathcal{S}$ is connected to node $n(i)\in \mathcal{N}$ in the transmission network. The supplier bids into the market by offering power at bid price $\alpha_{i,t}^p \in \mathbb{R}_+$ and offers available capacity $\bar{p}_{i,t} \in [0, \infty)$ for each $t \in \mathcal{T}$. We define $\mathcal{S}_n := \{i \in \mathcal{S} \, | \, n(i) = n\} \subseteq \mathcal{S}$ (set of suppliers connected to node $n$). The cleared allocation for supplier $i\in \mathcal{S}$ (load injected) are denoted as $p_{i,t}$ and must satisfy $p_{i,t} \in [0,\bar{p}_{i,t}]$.  We use $p$ to denote the collection of all cleared allocations. 
\\

Each consumer $j \in \mathcal{D}$ is connected to node $n(j)\in \mathcal{N}$. The consumer bids into the market by requesting power at bid price $\alpha_{j,t}^d \in \mathbb{R}_+$ and requests a maximum capacity $\bar{d}_{j,t} \in [0, \infty)$ for each $t\in\mathcal{T}$. We define $\mathcal{D}_n := \{j \in \mathcal{D} \, | \, n(j) = n\} \subseteq \mathcal{D}$ (set of consumers connected to node $n$). For simplicity, we assume that there is only one consumer at a given node ($\mathcal{D}_n$ are singletons). The cleared allocation for consumer $j\in \mathcal{D}$ (load withdrawn) is denoted as $d_{j,t}$ and must satisfy $d_{j,t} \in [0,\bar{d}_{j,t}]$.  We use $d$ to denote the collection of all cleared allocations. 
\\

The transmission owner owns the whole transmission network, where each (undirected) line $l \in \mathcal{L}$ is defined by its sending node $\mathrm{snd}(l) \in \mathcal{N}$ and receiving node $\mathrm{rec}(l) \in \mathcal{N}$. For the purpose of linearizing the market clearing problem (the details of which can be found in \cite{zhang2021electricity}), we decompose each line $l$ into two directed edges: $l^+ := (\mathrm{snd}(l),\mathrm{rec}(l))$ and $l^- := (\mathrm{rec}(l),\mathrm{snd}(l))$. The set of all directed edges from such decomposition is $\mathcal{K} := \cup_{l\in\mathcal{L}}\{l^+,l^-\}$. For each node $n \in \mathcal{N}$, we define its set of receiving lines $\mathcal{K}_n^{\textrm{rec}} := \{k \in \mathcal{K} \, | \, n=\textrm{rec}(k)\} \subseteq \mathcal{K}$ and its set of sending lines $\mathcal{K}_n^{\textrm{snd}} := \{k \in \mathcal{K} \, | \, n=\textrm{snd}(k)\}\subseteq \mathcal{K}$. Each line offers a bid price $\alpha^f_{k,t} \in \mathcal{R}_+$ and capacity $\bar{f}_{k,t} \in [0, \infty)$. Each cleared flow $f_{k,t}$ must satisfy the bounds $f_{k,t} \in [-\bar{f}_{k,t}, \bar{f}_{k,t}]$ and the collection $f$ must obey the direct-current (DC) power flow equations:
\begin{equation}
\label{eq:dc_flow}
f_{l^+,t} - f_{l^-,t} = B_l(\theta_{\textrm{snd}(l),t} - \theta_{\textrm{rec}(l),t}),
\end{equation}
where $B_l \in \mathbb{R}_+$ is the line susceptance and $\theta_n \in \mathbb{R}$ is the phase angle at node $n \in \mathcal{N}$. The DC power flow model is a linear model and requires small phase angle differences across transmission lines $\theta_{\textrm{snd}(l),t} - \theta_{\textrm{rec}(l),t} \in [-\Delta\bar{\theta'}_{l,t},\Delta\bar{\theta'}_{l,t}]$. The limits on phase angle differences and the capacity constraints for flows can be captured as:
\begin{equation}
\label{eq:angle_diff_cap}
-\Delta\bar{\theta}_{l,t} \leq \theta_{\textrm{snd}(l),t} - \theta_{\textrm{rec}(l),t} \leq \Delta\bar{\theta}_{l,t}
\end{equation}
where $\Delta \bar{\theta}_{l,t} := \min\{\bar{f}_{l,t}/B_{l,t}, \Delta \bar{\theta'}_{l,t} \}$. 
\\

\begin{figure}[h!]
    \centering
    \includegraphics[width=0.65\textwidth]{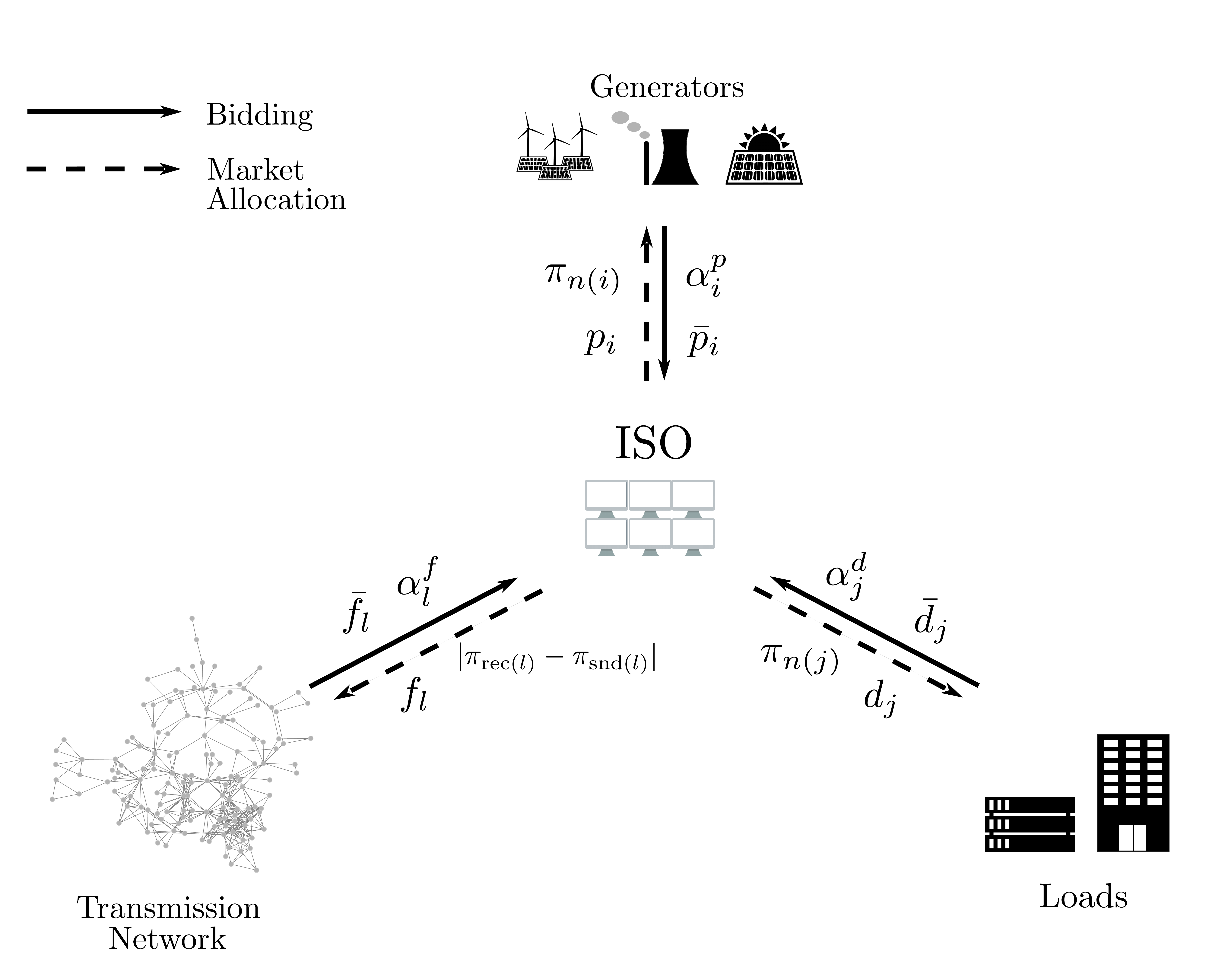}
    \caption{\small Sketch of base market clearing framework. }
    \label{fig:market}
\end{figure}

We use $\pi_{n,t}\in \mathbb{R}_+$ to represent the cleared price at the space-time node $(n,t) \in \mathcal{N} \times \mathcal{T}$. The collection of cleared prices is denoted as $\pi$; these are also known as nodal prices or LMPs and are used to charge/remunerate market players. We observe that, in a typical market, suppliers and transmission owners {\em offer a service} to the grid, while consumers {\em request a service} from the grid. Making this (rather) obvious distinction is important because we will see later that there are a couple ways in which we can capture ESR in a market clearing framework, which differ in whether ESR is treated as a pure service provider or a {\em prosumer} (service provider and consumer simultaneously). This standard market clearing process is illustrated in Figure \ref{fig:market}.

\subsection{Base Electricity Market Formulation}
We begin our discussion with a space-time market clearing formulation (with no ESR systems):
\begin{subequations}
\label{opt:standard}
\begin{align}
& \underset{d, p,f,\theta}{\text{min}} & & \sum_{t \in \mathcal{T}} \Big(\sum_{i \in \mathcal{S}} \alpha_{i,t}^p p_{i,t} + \sum_{k \in \mathcal{K}} \alpha^f_{k,t} f_{k,t} - \sum_{j\in\mathcal{D}} \alpha_{j,t}^d d_{j,t} \Big) \label{opt:standard_obj} \\ 
& {\text{s.t.}} & &  \sum_{k \in \mathcal{K}_n^{\textrm{rec}}} f_{k,t} + \sum_{i \in \mathcal{S}_n} p_{i,t} = \sum_{k \in \mathcal{K}_n^{\textrm{snd}}} f_{k,t} + \sum_{j \in \mathcal{D}_n} d_{j,t}, \quad (\pi_{n,t}) \quad  n \in \mathcal{N}, t \in \mathcal{T} \label{opt:standard_balance} \\ 
&&&  f_{l^+,t} - f_{l^-,t} = B_l(\theta_{\textrm{snd}(l),t} - \theta_{\textrm{rec}(l),t}), \quad l \in \mathcal{L}, t \in \mathcal{T} \label{opt:standard_flow} \\
&&& (d,p,\theta) \in \mathcal{C} \label{opt:standard_cap}
\end{align}
\end{subequations}
where $\mathcal{C} := \mathcal{C}^d \times \mathcal{C}^p \times \mathcal{C}^{\theta}$ captures the capacity constraints for all variables:
\begin{subequations}
\begin{gather}
\mathcal{C}_d := \{d\,| \, d_{j,t} \in [0, \bar{d}_{j,t}]\,\, \forall \,\, j \in \mathcal{D}, \, t \in \mathcal{T}\} \\
\mathcal{C}_p := \{p\,| \, p_{i,t} \in [0, \bar{p}_{i,t}]\,\, \forall \,\, i \in \mathcal{S}, \, t \in \mathcal{T}\} \\
\mathcal{C}_\theta := \{\theta \, | \, \theta_{\textrm{rec}(k),t} - \theta_{\textrm{snd}(k),t} \in [-\Delta\bar{\theta}_{k,t},\Delta\bar{\theta}_{k,t}] \,\, \forall \,\, k \in \mathcal{K}, t\in \mathcal{T}\}
\end{gather}
\end{subequations}
The objective function \eqref{opt:standard_obj} is known as the negative {\em social surplus} or {\em total welfare}, which captures the value of the demand served (to be maximized) and the cost of total supply and transmission services (to be minimized). The transmission cost is typically not included in the market clearing literature; this cost is included here to highlight an important analogy between transmission costs and power-shifting costs of ESR systems (to be discussed later). Constraint \eqref{opt:standard_balance} is the power balance constraint at each node $n$ (Kirchhoff's current law). Constraint \eqref{opt:standard_flow} is the DC power flow equation for each line $l$. 
\\

The solution of \eqref{opt:standard} gives the {\em primal} allocations $(p,d,f)$ and the {\em dual} allocations $\pi$. The dual allocations are the optimal solutions of the dual variables associated with the power balance constraints \eqref{opt:standard_balance}. These dual allocations are used as locational marginal prices (LMPs) of electricity that clear the market. This market clearing process is illustrated in Figure \ref{fig:market}. Most often, the prices are different across different locations and times (thus giving rise fo space-time volatility). We use $(p,d,f,\pi)$ to denote the primal-dual allocation obtained from the solution of the clearing formulation. Under this market, each supplier $i$ is cleared at the price $\pi_{n(i),t}$ for each unit of power provided at time $t$, and each consumer $j$ pays at the price $\pi_{n(j), t}$ for each unit of power consumed at time $t$. Each transmission line $l$ is remunerated by the price difference $|\pi_{\textrm{snd}(l),t} - \pi_{\textrm{rec}(l),t}|$ at time $t$. The profit function for each participant at each time is defined as follows:
\begin{subequations}
\label{eq:standard_profit}
\begin{gather}
\phi^p_{i,t} := (\pi_{n(i), t} - \alpha^p_{i,t}) p_{i,t} \\
\phi^d_{j,t} := (\alpha^d_{j,t} - \pi_{n(j), t}) d_{j,t} \\
\phi^f_{k,t} := (\pi_{\textrm{rec}(l),t} - \pi_{\textrm{snd}(l),t} - \alpha^f_k) f_k
\end{gather}
\end{subequations}

\subsection{Base Market Formulation with ESR}
We now extend the space-time market formulation to capture ESR systems as a new type of market participants. To do that, the market is extended to consider a set of ESR systems $\mathcal{B}$. Each ESR $b\in\mathcal{B}$ is connected to the power grid at node $n(b) \in \mathcal{N}$. The ESR bids into the market by offering charging service at bid price $\alpha^{sc}_{b,t}$ and discharging service at bid price $\alpha^{sd}_{b,t}$ for each time $t\in\mathcal{T}$. 
\\

The clearing formulation for markets with ESR participants is presented below:
\begin{subequations}
\label{opt:market_esr}
\begin{align}
& \underset{d, p,f,\theta, p^c, p^d}{\text{min}} & & \sum_{t \in \mathcal{T}} \Big(\sum_{i \in \mathcal{S}} \alpha_{i,t}^p p_{i,t} + \sum_{k \in \mathcal{K}} \alpha^f_{k,t} f_{k,t} - \sum_{j\in\mathcal{D}} \alpha_{j,t}^d d_{j,t} + \sum_{b\in\mathcal{B}}\alpha^{sc}_{b,t} p^c_{b,t} + \sum_{b\in\mathcal{B}}\alpha^{sd}_{b,t} p^d_{b,t}\Big) \label{opt:market_esr_obj} \\
& {\text{s.t.}} & & \sum_{k \in \mathcal{K}_n^{\textrm{rec}}} f_{k,t} + \sum_{i \in \mathcal{S}_n} p_{i,t} + \sum_{b\in\mathcal{B}_n} p^d_{b,t} = \sum_{k \in \mathcal{K}_n^{\textrm{snd}}} f_{k,t} + \sum_{j \in \mathcal{D}_n} d_{j,t} + \sum_{b\in\mathcal{B}_n} p^c_{b,t}, \, (\pi_{n,t}) \,  n \in \mathcal{N}, t \in \mathcal{T} \label{opt:market_esr_bal} \\ 
&&&  f_{l^+,t} - f_{l^-,t} = B_l(\theta_{\textrm{snd}(l),t} - \theta_{\textrm{rec}(l),t}), \quad l \in \mathcal{L}, t \in \mathcal{T} \label{opt:market_esr_flow} \\
&&& \Delta\underline{s}_{b,t} \leq \eta^c_b \sum_{t' = 1}^t p^c_{b,t'} - \frac{1}{\eta^d_b} \sum_{t' = 1}^t p^d_{b,t'} \leq \Delta\bar{s}_b, \quad b\in\mathcal{B}, t\in\mathcal{T} \label{opt:market_esr_energy_cap} \\
&&& p^c_{b,t} + p^d_{b,t} \leq \bar{p}_b, \quad b\in\mathcal{B}, t\in\mathcal{T} \label{opt:market_esr_power_cap} \\
&&& p^c_{b,t}, p^d_{b,t} \geq 0, \quad b\in\mathcal{B}, t\in\mathcal{T} \label{opt:market_esr_power_nonneg} \\
&&& (d,p,\theta) \in \mathcal{C} \label{opt:market_esr_c}
\end{align}
\end{subequations}
The social surplus \eqref{opt:market_esr_obj} captures the charging and discharging costs of ESR over the entire time horizon. The nodal power balances \eqref{opt:market_esr_bal} capture charging and discharging power of ESR. Constraints \eqref{opt:market_esr_energy_cap}--\eqref{opt:market_esr_power_nonneg} capture the operational constraints of each ESR participant. Note that \eqref{opt:market_esr_energy_cap}--\eqref{opt:market_esr_power_nonneg} are equivalent to \eqref{eq:storage_nos} except for \eqref{opt:market_esr_power_nonneg}, where the complementarity constraint in \eqref{eq:storage_nos_perp} is relaxed. This means it is possible for the market to deliver power allocations and prices that result in infeasible operation for ESRs. However, we will show that the optimal solutions of \eqref{opt:market_esr} satisfy the complementarity constraints \eqref{eq:storage_nos_perp} under non-negative prices. 
\\

The objective function \eqref{opt:market_esr_obj} is different from most market clearing formulations that capture ESR in literature, where the charging terms of ESR are not treated as a cost but a consumer surplus (i.e., they are assigned the same sign as the load term). There are a couple of main disadvantages with such formulation. From the modeling perspective, treating the charging terms as a consumer surplus implies that ESR participants benefit only from the charging actions in the market because they are getting electricity. This is not consistent with reality because energy arbitrage requires both charging and discharging actions, so charging without discharging does not add value to ESR. From the computational perspective, assigning opposite signs for charging and discharging leads to optimal solutions with simultaneous charging and discharging, which makes the complementarity constraints necessary. 
\\

We now establish the market properties of \eqref{opt:market_esr} to demonstrate how to remunerate ESR participants. The partial Lagrange function is:
\begin{equation}
\begin{aligned}
    L(d,p,f,\theta,p^c,p^d,\pi) &= \sum_{t \in \mathcal{T}} \Big(\sum_{i \in \mathcal{S}} \alpha_{i,t}^p p_{i,t} + \sum_{k \in \mathcal{K}} \alpha^f_{k,t} f_{k,t} - \sum_{j\in\mathcal{D}} \alpha_{j,t}^d d_{j,t} + \sum_{b\in\mathcal{B}}\alpha^{sc}_{b,t} p^c_{b,t} + \sum_{b\in\mathcal{B}}\alpha^{sd}_{b,t} p^d_{b,t}\Big)\\
    &\quad -\sum_{n\in\mathcal{N}, t\in\mathcal{T}} \pi_{n,t} \big( \sum_{k \in \mathcal{K}_n^{\textrm{rec}}} f_{k,t} + \sum_{i \in \mathcal{S}_n} p_{i,t} + \sum_{b\in\mathcal{B}_n} p^d_{b,t} - \sum_{k \in \mathcal{K}_n^{\textrm{snd}}} f_{k,t} - \sum_{j \in \mathcal{D}_n} d_{j,t} - \sum_{b\in\mathcal{B}_n} p^c_{b,t} \big) \\
    &= -\sum_{t\in\mathcal{T}}\Bigg\{ \sum_{j \in \mathcal{N}_d} (\alpha^d_{j,t} - \pi_{n(j)})d_{j,t} + \sum_{i \in \mathcal{S}}(\pi_{n(i),t} - \alpha_i^p)p_{i,t} \\
    &\qquad+ \sum_{k \in \mathcal{K}}(\pi_{\textrm{rec}(k)} + \pi_{\textrm{snd}(k)} + \alpha^f_{k,t}) f_{k,t} + \sum_{b \in \mathcal{B}} \bigg[ (\pi_{b,t} - \alpha^{sd}_{b,t} )p^d_{b,t} - (\pi_{b,t} + \alpha^{sc}_{b,t})p^c_{b,t} \bigg]  \Bigg\}
\end{aligned}
\end{equation}
and the Lagrange dual problem is:
\begin{subequations}
\begin{align}
\max_\pi \mathcal{D}(\pi) := & \underset{d, p,f,\theta, p^c, p^d}{\text{min}} & & L(d,p,f,\theta,p^c,p^d,\pi) \\
& {\text{s.t.}} &&  f_{l^+,t} - f_{l^-,t} = B_l(\theta_{\textrm{snd}(l),t} - \theta_{\textrm{rec}(l),t}), \quad l \in \mathcal{L}, t \in \mathcal{T} \\
&&& \underline{s}_b - s_{b,0} \leq \eta^c_b \sum_{t' = 1}^t p^c_{b,t'} - \frac{1}{\eta^d_b} \sum_{t' = 1}^t p^d_{b,t'} \leq \bar{s}_b - s_{b,0}, \quad b\in\mathcal{B}, t\in\mathcal{T} \\
&&& 0 \leq p^c_{b,t} + p^d_{b,t} \leq \bar{p}_b, \quad b\in\mathcal{B}, t\in\mathcal{T} \\
&&& p^c_{b,t}, p^d_{b,t} \geq 0, \quad b\in\mathcal{B}, t\in\mathcal{T} \\
&&& \eta^c_b \sum_{t' = 1}^T p^c_{b,t'} - \frac{1}{\eta^d_b} \sum_{t' = 1}^T p^d_{b,t'} \geq 0, \quad b\in\mathcal{B} \\
&&& (d,p,\theta) \in \mathcal{C}
\end{align}
\end{subequations}
The Lagrange dual function $\mathcal{D}(\pi)$ can be decomposed to individual profit maximization problems. Here we present the ESR profit maximization problem for each $b\in\mathcal{B}$:
\begin{subequations}
\label{opt:esr_profit_max}
\begin{align}
& \underset{p^c_b, p^d_b}{\text{max}} & & \sum_{t \in \mathcal{T}} (\pi_{b,t} - \alpha^{sd}_{b,t})p^d_{b,t} - (\pi_{b,t} + \alpha^{sc}_{b,t})p^c_{b,t} \label{opt:esr_profit_max_obj} \\
& {\text{s.t.}} && \underline{s}_b - s_{b,0} \leq \eta^c_b \sum_{t' = 1}^t p^c_{b,t'} - \frac{1}{\eta^d_b} \sum_{t' = 1}^t p^d_{b,t'} \leq \bar{s}_b - s_{b,0}, \quad t\in\mathcal{T} \label{opt:esr_profit_max_energy_cap} \\
&&& 0 \leq p^c_{b,t} + p^d_{b,t} \leq \bar{p}_b, \quad t\in\mathcal{T} \\
&&& p^c_{b,t}, p^d_{b,t} \geq 0, \quad t\in\mathcal{T} \\
&&& \eta^c_b \sum_{t' = 1}^T p^c_{b,t'} - \frac{1}{\eta^d_b} \sum_{t' = 1}^T p^d_{b,t'} \geq 0 \label{opt:esr_profit_max_final}
\end{align}
\end{subequations}
Problem \eqref{opt:esr_profit_max} gives some intuition on why the optimal dispatch of ESR satisfies the complementarity constraint under non-negative prices. We observe that for any non-negative value of $\pi$, the objective function will not incentivize charging and discharging at the same time. Specifically, at each time interval $t\in\mathcal{T}$, discharging is incentivized when $\pi_{b,t} \geq \alpha^{sd}_{b,t} > 0$, and charging is incentivized when $\pi_{b,t} \leq -\alpha^{sc}_{b,t} < 0$. Note that under non-negative prices, the latter condition will never occur. However, this does not mean ESR will not charge at all: the lower bound on state of charge (first inequality of \eqref{opt:esr_profit_max_energy_cap}) will enforce the ESR to charge enough power at times with lower prices. This also implies that the ESR will not charge more than necessary with non-negative electricity prices. This is consistent with the notion that charging is part of the service that ESR offers to transport electricity in time. 
\\

With the profit maximization problem \eqref{opt:esr_profit_max}, we are able to establish the complementarity condition rigorously. Let $(d^*,p^*,f^*,\theta^*,p^{c*}, p^{d*}), \pi^*$ denote the optimal primal-dual solution of \eqref{opt:market_esr}. A sufficient condition for complementarity is as follows.
\begin{theorem}
\label{thm:complementarity_condition}
For any $b \in \mathcal{B}, t \in \mathcal{T}$, we have that $p^{c*}_{b,t} \cdot p^{d*}_{b,t} = 0$ if
\begin{equation}
\label{eq:complementarity_condition}
\pi^*_{b,t} \geq -\frac{1}{1-\eta_b}(\eta_b \alpha^{sd}_{b,t} + \alpha^{sc}_{b,t})    
\end{equation}
\end{theorem}
\begin{proof}
See Appendix.
\end{proof}
Condition \eqref{eq:complementarity_condition} gives a lower bound for prices so that ESR will not benefit from charging and discharging simultaneously. The bound will be violated by negative prices, where there is local excess power and ESR is encouraged to charge and discharge simultaneously in order to consume these excess power without violating the SOC constraints in the ESR itself. In practice, various inflexible components of power systems may lead to such negative prices, such as inflexible operations of renewable energy and ramp limits of generators. Note that, if the ESR is ideal (efficiency of $\eta_b = 1$), the market formulation always recovers an optimal solution that satisfies complementarity (regardless of the prices). Also, we note that \eqref{eq:complementarity_condition} is a sufficient but not necessary condition because other constraints (e.g. SOC bounds) may also prevent simultaneous charging and discharging despite negative prices. We will see how this will allow us to design an alternative market ensures charge/discharge complementarity without enforcing complementarity constraints explicitly (thus facilitating computational tractability).  

\subsection{Alternative Market Formulation with Robust Bound}

So far we have shown that simply relaxing the complementarity constraints may lead to allocations for ESRs that are not physically realizable. To tackle this issue, we propose an alternative formulation based on the robust battery dispatch formulation recently proposed in \cite{nazir2021guaranteeing}. The basic idea is to approximate the charge and discharge operations of ESR with a series of net-charge decisions $p^c_{b,t}-p^d_{b,t}$ with a net-charge efficiency $\frac{\eta^c_b}{\eta^d_b}$. Instead of including an exact upper bound for SOC as in \eqref{opt:market_esr_energy_cap}, we replace it with a conservative upper bound function as shown on the right-hand side below: 
\begin{equation}
\label{eq:soc_ub_function}
    \eta^c_b \sum_{t' = 1}^t p^c_{b,t'} - \frac{1}{\eta^d_b} \sum_{t' = 1}^t p^d_{b,t'} \leq  \frac{\eta^c_b}{\eta^d_b}\left(\sum_{t' = 1}^t p^c_{b,t'} - p^d_{b,t'}\right)
\end{equation}
This leads to the following market formulation.
\begin{subequations}
\label{opt:robust_market_esr}
\begin{align}
& \underset{d, p,f,\theta, p^c, p^d}{\text{min}} & & \sum_{t \in \mathcal{T}} \Big(\sum_{i \in \mathcal{S}} \alpha_{i,t}^p p_{i,t} + \sum_{k \in \mathcal{K}} \alpha^f_{k,t} f_{k,t} - \sum_{j\in\mathcal{D}} \alpha_{j,t}^d d_{j,t} + \sum_{b\in\mathcal{B}}\alpha^{sc}_{b,t} p^c_{b,t} + \sum_{b\in\mathcal{B}}\alpha^{sd}_{b,t} p^d_{b,t}\Big) \label{opt:robust_market_esr_obj} \\
& {\text{s.t.}} & & \sum_{k \in \mathcal{K}_n^{\textrm{rec}}} f_{k,t} + \sum_{i \in \mathcal{S}_n} p_{i,t} + \sum_{b\in\mathcal{B}_n} p^d_{b,t} = \sum_{k \in \mathcal{K}_n^{\textrm{snd}}} f_{k,t} + \sum_{j \in \mathcal{D}_n} d_{j,t} + \sum_{b\in\mathcal{B}_n} p^c_{b,t}, \, (\pi_{n,t}) \,  n \in \mathcal{N}, t \in \mathcal{T} \label{opt:robust_market_esr_bal} \\ 
&&&  f_{l^+,t} - f_{l^-,t} = B_l(\theta_{\textrm{snd}(l),t} - \theta_{\textrm{rec}(l),t}), \quad l \in \mathcal{L}, t \in \mathcal{T} \label{opt:robust_market_esr_flow} \\
&&& \eta^c_b \sum_{t' = 1}^t p^c_{b,t'} - \frac{1}{\eta^d_b} \sum_{t' = 1}^t p^d_{b,t'} \geq \Delta\underline{s}_{b,t}, \quad b\in\mathcal{B}, t\in\mathcal{T} \label{opt:robust_market_esr_lb} \\
&&& \frac{\eta^c_b}{\eta^d_b} \left( \sum_{t' = 1}^t p^c_{b,t'} - p^d_{b,t'} \right)\leq \Delta\bar{s}_b, \quad b\in\mathcal{B}, t\in\mathcal{T} \label{opt:robust_market_esr_ub} \\
&&& p^c_{b,t} + p^d_{b,t} \leq \bar{p}_b, \quad b\in\mathcal{B}, t\in\mathcal{T} \label{opt:robust_market_esr_power_cap} \\
&&& p^c_{b,t}, p^d_{b,t} \geq 0, \quad b\in\mathcal{B}, t\in\mathcal{T} \label{opt:robust_market_esr_power_nonneg} \\
&&& (d,p,\theta) \in \mathcal{C} \label{opt:robust_market_esr_c}
\end{align}
\end{subequations}
Note that the only difference between \eqref{opt:market_esr} and \eqref{opt:robust_market_esr} is that the robust SOC upper bound \eqref{opt:robust_market_esr_ub} replaces the exact SOC upper bound in \eqref{opt:market_esr_energy_cap}. Also, the inequality \eqref{eq:soc_ub_function} implies that the feasible region of \eqref{opt:market_esr} is a subset of the feasible region of \eqref{opt:robust_market_esr}. This means formulation \eqref{opt:robust_market_esr} compromises the optimal total welfare value in exchange for guaranteeing a feasible operation for ESRs.
\\

Because only one of the constraints for ESR is changed, we can easily obtain the profit maximization problem for each ESR using the same Lagrangian dual analysis as follows.
\begin{subequations}
\label{opt:robust_esr_profit_max}
\begin{align}
& \underset{p^c_b, p^d_b}{\text{max}} & & \sum_{t \in \mathcal{T}} (\pi_{b,t} - \alpha^{sd}_{b,t})p^d_{b,t} - (\pi_{b,t} + \alpha^{sc}_{b,t})p^c_{b,t} \label{opt:robust_esr_profit_max_obj} \\
& {\text{s.t.}} && \eta^c_b \sum_{t' = 1}^t p^c_{b,t'} - \frac{1}{\eta^d_b} \sum_{t' = 1}^t p^d_{b,t'} \geq \Delta \underline{s}_{b,t}, \quad t\in\mathcal{T} \label{opt:robust_esr_profit_lb} \\
&&& \frac{\eta^c_b}{\eta^d_b} \left(\sum_{t' = 1}^t p^c_{b,t'} - p^d_{b,t'}\right) \leq \Delta\bar{s}_b, \quad t\in\mathcal{T} \label{opt:robust_esr_profit_ub} \\
&&& p^c_{b,t} + p^d_{b,t} \leq \bar{p}_b, \quad t\in\mathcal{T} \\
&&& p^c_{b,t}, p^d_{b,t} \geq 0, \quad t\in\mathcal{T}
\end{align}
\end{subequations}
Let $(d^*,p^*,f^*,\theta^*,p^{c*}, p^{d*}), \pi^*$ denote the optimal primal-dual solution of \eqref{opt:robust_market_esr}.
\begin{theorem}
\label{thm:robust_complementarity}
Formulation \eqref{opt:robust_market_esr} delivers allocations that satisfy the complementarity $p^{c*}_{b,t} \cdot p^{d*}_{b,t} = 0$ for any $b \in \mathcal{B}, t \in \mathcal{T}$.
\end{theorem}
\begin{proof}
See Appendix.
\end{proof}

The previous theorem is significant because it avoids the need to explicitly incorporate complementarity constraints in the market clearing formulation. Such constraints are difficult to handle computationally, as they introduce nonconvexity. 

\subsection{Market Formulation with Virtual Links}

We have seen that price differences in time provide economic incentives for ESRs. Under formulation \eqref{opt:market_esr}, each ESR participant make a profit via energy arbitrage at the connected nodes. It is widely believed that ESRs can offer flexibility for electricity markets, and such flexibility helps mitigate price volatility (risk). However, none of these advantages is obvious from formulation \eqref{opt:market_esr}. This motivates us to propose a mathematically-equivalent formulation that models flexibility of ESR with the notion of virtual links; specifically, the charging and discharging of power can be seen as a temporal transfer (transport) of power from the charging time to discharging time. 
\\

In the alternative formulation, we view the operations of ESRs in a different way. Instead of modeling their operations as a series of charging and discharging decisions over time, we model the operations in terms of how small packages of energy are moved between the grid and ESRs. Specifically, we break down the operations of ESRs into the following three categories:
\begin{enumerate}
    \item {\em Net-charging:} buying an amount of electricity from the market at a time period and storing it for the rest of the period.
    \item {\em Net-discharging:} selling an amount of electricity to the market at a time period that will not be replaced by electricity purchase later.
    \item {\em Energy transfer:} moving certain amount of energy from one time to another time. This captures charging/discharging certain amount of electricity at one time and discharging/charging it later.
\end{enumerate}
Note that energy transfer operations have no effect on the net change of state of charge for the ESR over the whole time period of market clearing, while net-charging and net-discharging operations do have such effects. 

\begin{figure}[!htp]
    \centering
    \includegraphics[width=0.8\textwidth,trim={3cm 0 3cm 0},clip]{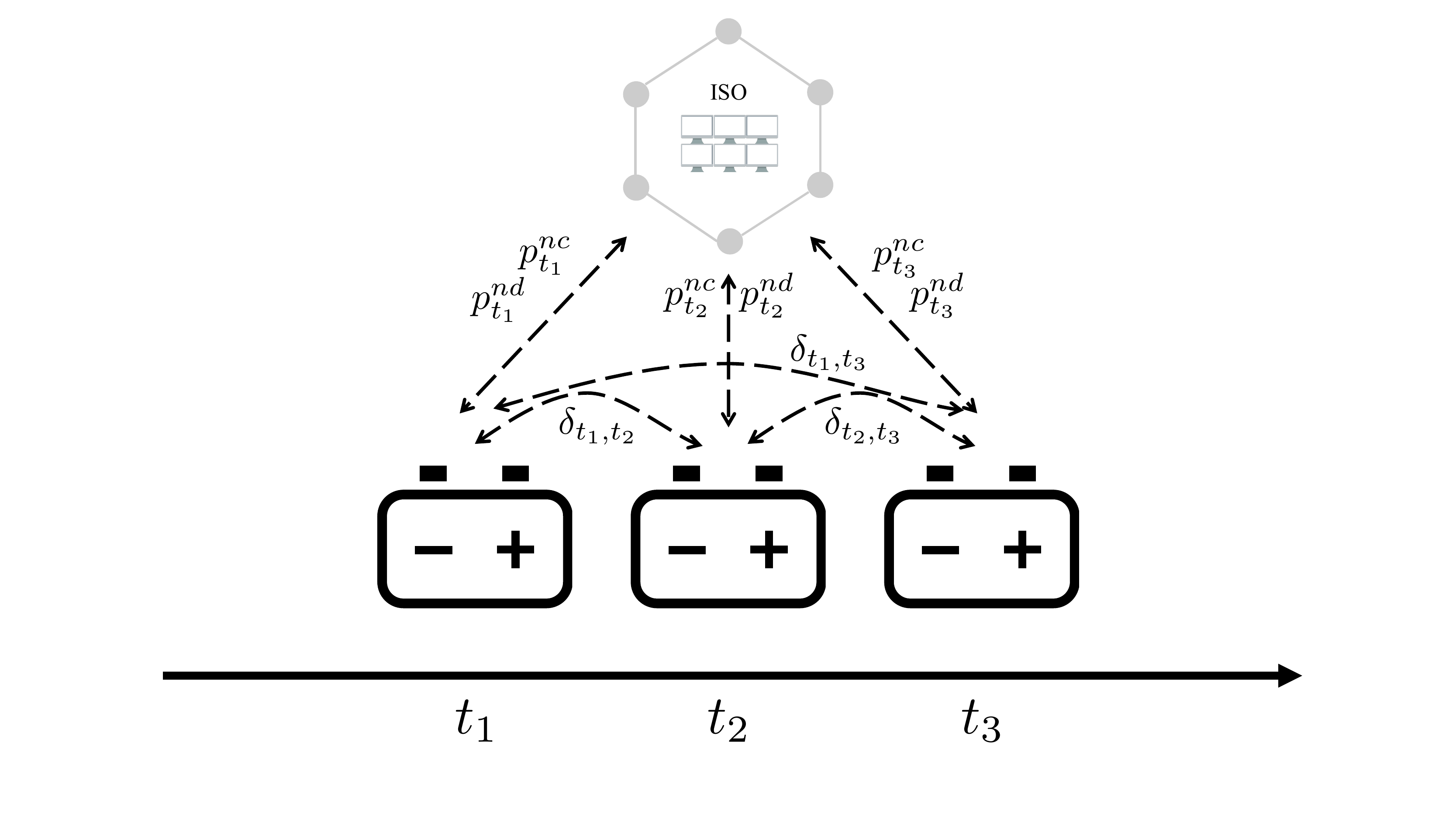}
    \caption{\small Alternative formulation for ESR systems using virtual links (ESR index $b$ is omitted in notations). }
    \label{fig:storage_vl}
\end{figure}

The ESR modeling of the alternative formulation is illustrated in Figure \ref{fig:storage_vl}. We define decision variables $p^{nc}_{b,t} \in \mathbb{R}_+, p^{nc}_{b,t} \in \mathbb{R}_+$ to capture the net-charging and net-discharging operations of each ESR $b \in \mathcal{B}$ at time $t \in \mathcal{T}$. To model energy transfer, we extend the concept of virtual links proposed in the market design work in \cite{zhang2021electricity}. There, virtual links are proposed to capture non-physical lossless shift of electricity loads enabled by geo-distributed computing infrastructure. Here, we extend this concept to capture non-physical (meaning not via the physical transmission grid) energy transfer in time enabled by ESRs. 
\\

Let $\mathcal{V}$ be the set of all virtual links; each virtual link $v = (b(v), t_c(v), t_d(v)) \in \mathcal{V}$ has an associated ESR $b(v)$, charging time $t_c(v)$, and discharging time $t_d(v)$. We require that $t_c(v) \neq t_d(v)$ for all $v \in \mathcal{V}$. Implicitly, each virtual link $v$ is also associated with a location $n(v) := n(b(v))$ from the associated ESR. We define $\mathcal{V}_{n,t}^{\textrm{in}} := \{v \in \mathcal{V} \, | \, t_d(v) = t, n(v) = n \}$, $\mathcal{V}_{n,t}^{\textrm{out}} := \{v \in \mathcal{V} \, | \, t_c(v) = t, n(v) = n\}$ as the set of discharging and charging virtual links at space-time node $(n,t)$, respectively. These sets form two partitions of virtual links on the dimension of space-time nodes: $\mathcal{V} = \cup_{(n,t) \in\mathcal{N}\times\mathcal{T}} \mathcal{V}_{n,t}^{\textrm{in}} = \cup_{(n,t) \in\mathcal{N}\times\mathcal{T}} \mathcal{V}_{n,t}^{\textrm{out}}$. We also define $\mathcal{V}_{b,t}^{\textrm{in}} := \{v \in \mathcal{V} \, | \, t_d(v) = t, b(v) = b \}$, $\mathcal{V}_{b,t}^{\textrm{out}} := \{v \in \mathcal{V} \, | \, t_c(v) = t, b(v) = b\}$ as the set of discharging and charging virtual links for ESR $b$ at time $t$, respectively. These sets further form partitions for the sets $\mathcal{V}_{n,t}^{\textrm{in}}$ and $\mathcal{V}_{n,t}^{\textrm{out}}$: $\mathcal{V}_{n,t}^{\textrm{in}} = \cup_{b\in\mathcal{B}_n} \mathcal{V}_{b,t}^{\textrm{in}}$, $\mathcal{V}_{n,t}^{\textrm{out}} = \cup_{b\in\mathcal{B}_n} \mathcal{V}_{b,t}^{\textrm{out}}$. Each virtual link $v$ is also associated with a bid price for the power shift.
\\

The cleared power shifts (virtual flows) are defined as $\delta_v \in \mathbb{R}_+$. Note that shifting power via ESR systems incurs loss of power due to charging and/or discharging inefficiency (i.e., $\eta_c, \eta_d < 1$). This means the amount of power discharged may be strictly less than the amount of power charged for each virtual link. For consistency, we define $\delta_v$ as the amount of power charged by virtual link $v$ at time $t_c(v)$; the amount of power discharged at time $t_d(v)$ by $v$ can be calculated as $\eta_{b(v)}\delta_v$, where $\eta_{b(v)} := \eta_{b(v)}^c\cdot\eta_{b(v)}^d$ is the aggregated (round-trip) efficiency of a pair of charging and discharging actions for ESR $b(v)$. Now we can compute the charging and discharging power $p^c_{b,t},p^d_{b,t}$ as follows:
\begin{subequations}
\label{eq:operations_mapping}
\begin{gather}
    p_{b,t}^c = \sum_{v\in\mathcal{V}^{\textrm{out}}_{b,t}}\delta_v + p_{b,t}^{nc}\\
    p_{b,t}^d = \sum_{v\in\mathcal{V}^{\textrm{in}}_{b,t}}\eta_{b(v)} \delta_v + p_{b,t}^{nd}
\end{gather}
\end{subequations}
Then the total amount of power charged and discharged by all ESRs at space-time node $(n,t)$ can be expressed in terms of $\delta$ as follows:
\begin{subequations}
\begin{gather}
    \sum_{b\in\mathcal{B}_n} p^c_{b,t} = \sum_{{b\in\mathcal{B}_n}}\left[\sum_{v\in\mathcal{V}^{\textrm{out}}_{b,t}} \delta_v + p^{nc}_{b,t}\right]= \sum_{v\in\mathcal{V}^{\textrm{out}}_{n,t}} \delta_v + \sum_{b\in\mathcal{B}_n} p^{nc}_{b,t}\\
    \sum_{b\in\mathcal{B}_n} p^d_{b,t} = \sum_{{b\in\mathcal{B}_n}}\left[\sum_{v\in\mathcal{V}^{\textrm{in}}_{b,t}}\eta_{b(v)} \delta_v + p^{nd}_{b,t}\right]= \sum_{v\in\mathcal{V}^{\textrm{in}}_{n,t}} \eta_{b(v)} \delta_v + \sum_{b\in\mathcal{B}_n} p^{nd}_{b,t}
\end{gather}
\end{subequations}
The clearing formulation with virtual links is:
\begin{subequations}
\label{opt:vl}
\begin{align}
& \underset{d, p,f,\theta,\delta}{\text{min}} & & \sum_{t \in \mathcal{T}} \Big(\sum_{i \in \mathcal{S}} \alpha_{i,t}^p p_{i,t} + \sum_{k \in \mathcal{K}} \alpha^f_{k,t} f_{k,t} - \sum_{j\in\mathcal{D}} \alpha_{j,t}^d d_{j,t} \Big) + \sum_{v\in\mathcal{V}}\alpha^\delta_v \delta_v \label{opt:vl_obj}\\ 
& {\text{s.t.}} & & \sum_{k \in \mathcal{K}_n^{\textrm{rec}}} f_{k,t} + \sum_{i \in \mathcal{S}_n} p_{i,t} + \sum_{v\in\mathcal{V}_{n,t}^{\textrm{in}}} \eta_{b(v)}\delta_v + \sum_{b\in\mathcal{B}_n} p^{nd}_{b,t} \nonumber \\
&&& \quad\quad\quad = \sum_{k \in \mathcal{K}_n^{\textrm{snd}}} f_{k,t} + \sum_{j \in \mathcal{D}_n} d_{j,t} + \sum_{v\in\mathcal{V}_{n,t}^{\textrm{out}}} \delta_v + \sum_{b\in\mathcal{B}_n} p^{nc}_{b,t}, \, (\pi_{n,t}) \,  n \in \mathcal{N}, t \in \mathcal{T} \label{opt:vl_bal}\\ 
&&&  f_{l^+,t} - f_{l^-,t} = B_l(\theta_{\textrm{snd}(l),t} - \theta_{\textrm{rec}(l),t}), \quad l \in \mathcal{L}, t \in \mathcal{T} \label{opt:vl_flow}\\
&&& \eta^c_b\sum_{t'=1}^t \left[ \sum_{v\in\mathcal{V}^{\textrm{out}}_{b,t'}} \delta_v - \sum_{v\in\mathcal{V}^{\textrm{in}}_{b,t'}} \delta_v \right] \geq \Delta\underline{s}_{b,t} + \frac{1}{\eta^d_b} \sum_{t'=1}^t p^{nd}_{b,t'} , \quad b\in\mathcal{B}, t\in\mathcal{T} \label{opt:vl_soc_lb}\\ 
&&& \frac{\eta^c_b}{\eta^d_b} \sum_{t'=1}^t \left[ \sum_{v\in\mathcal{V}^{\textrm{out}}_{b,t'}} \delta_v - \sum_{v\in\mathcal{V}^{\textrm{in}}_{b,t'}} \eta_b\delta_v \right] \leq \Delta\bar{s}_b - \eta^c_b \sum_{t'=1}^t p^{nc}_{b,t'}, \quad b\in\mathcal{B}, t\in\mathcal{T} \label{opt:vl_soc_ub}\\
&&& p^{nc}_{b,t} + p^{nd}_{b,t} +  \sum_{v\in\mathcal{V}^{\textrm{out}}_{b,t}} \delta_v + \sum_{v\in\mathcal{V}^{\textrm{in}}_{b,t}}\eta_b \delta_v \leq \bar{p}_b, \quad b\in\mathcal{B}, t\in\mathcal{T} \label{opt:vl_power_cap}\\
&&& \delta \geq 0 \label{opt:vl_nonneg} \\
&&& (d,p,\theta) \in \mathcal{C} \label{opt:vl_C}
\end{align}
\end{subequations}

\begin{theorem}
\label{thm:vl_equivalence}
Let $\alpha^\delta_v = \alpha^{sc}_{b(v),t_c(v)} + \eta_{b(v)}\alpha^{sd}_{b(v),t_d(v)}$ for each $v \in \mathcal{V}$. Then, formulations \eqref{opt:vl} and \eqref{opt:robust_market_esr} are equivalent. 
\end{theorem}
\begin{proof}
See Appendix.
\end{proof}

\begin{figure}[h!]
    \centering
    \includegraphics[width=0.65\textwidth]{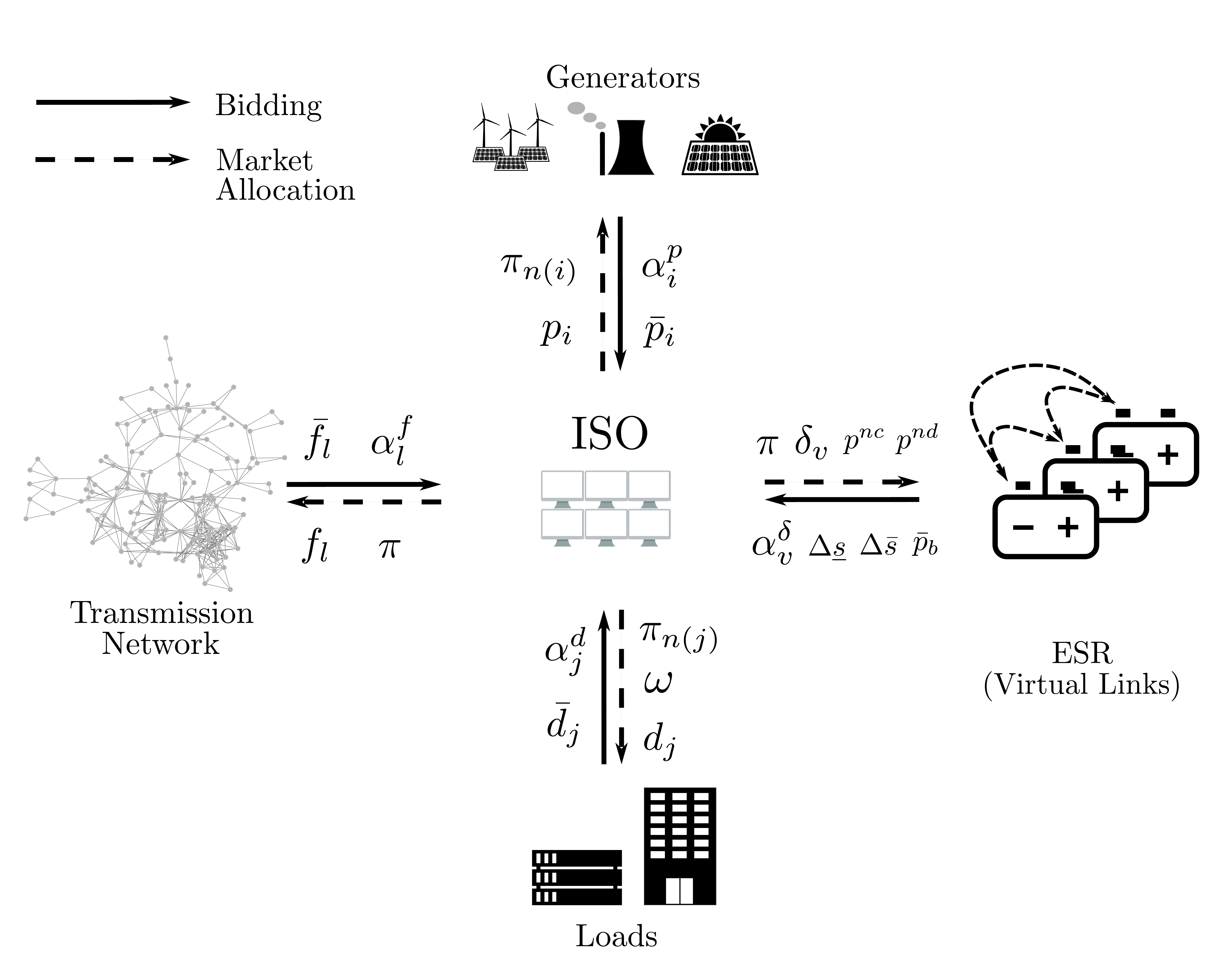}
    \caption{\small Market clearing framework with ESR using virtual links. }
    \label{fig:market_vl}
\end{figure}

Theorem \ref{thm:vl_equivalence} shows the equivalence between the market formulations when each virtual link is assigned a proper bid price. The bid price assigned for each virtual link covers the cost for the corresponding charging and discharging operations. Also, the bid price is a function of the round-trip efficiency to account for lost power in the transfer process. Figure \ref{fig:market_vl} illustrates the market clearing process with virtual links modeling the flexibility of ESR systems. 
\\

A key advantage of the virtual link-based framework, compared to \eqref{opt:market_esr} or \eqref{opt:robust_market_esr}, is the ability to allow for more sophisticated bidding by ESRs. In Theorem \ref{thm:vl_equivalence}, the equivalence between the formulations require that each virtual link has the same bid price. More generally, however, each ESR is allowed to submit different bid prices for its virtual links. For instance, ESRs can submit a higher bid cost for virtual links that span a longer period of time, as the energy transferred via those virtual links will occupy the charge storage space for longer compared to those transferred via other virtual links. This is not easily viable in \eqref{opt:robust_market_esr}, as bidding is made for charging and discharging operations as a whole. 

Under the proposed market clearing framework, each ESR makes a payment for every unit of electricity it purchases, and gets remunerated for every unit of electricity sold to the market, as well as every unit of energy transfer via its virtual links. The total remuneration for each ESR is:
\begin{equation}
\label{eq:remuneration_vl}
\sum_{v\in\mathcal{V}_b} \pi_v^d(\eta_{b}\delta_v) +  \sum_{t \in \mathcal{T}} \pi_{b,t}p^{nd}_{b,t} - \sum_{v\in\mathcal{V}_b} \delta_v \pi_v^c - \sum_{t \in \mathcal{T}} \pi_{b,t} p^{nc}_{b,t}= \sum_{v\in\mathcal{V}_b}( \eta_{b}\pi_v^d - \pi^c_v)\delta_v + \sum_{t \in \mathcal{T}} \pi_{b,t}(p^{nd}_{b,t} -p^{nc}_{b,t})
\end{equation}
where $\pi^d_v := \pi_{n(b(v)), t_d(v)}$, $\pi^c_v := \pi_{n(b(v)), t_c(v)}$, $\delta_b := \{\delta_v\}_{v\in\mathcal{V}_b}$, $p^{nc}_b := \{p^{nc}_{b,t}\}_{t\in\mathcal{T}}$, $p^{nd}_b := \{p^{nd}_{b,t}\}_{t\in\mathcal{T}}$. This reveals the different elements of the service remuneration.
\\

The profit maximization problem for each ESR $b$ under the virtual link framework is:
\begin{subequations}
\label{opt:esr_profit_max_vl}
\begin{align}
& \underset{\delta_b, p^{nc}_b, p^{nd}_b}{\text{max}} & & \sum_{v\in\mathcal{V}_b} (\eta_{b} \pi_v^d - \pi^c_v - \alpha^\delta_v )\delta_v + \sum_{t \in \mathcal{T}}\left[ (\pi_{b,t} - \alpha^{sd}_{b,t})p^{nd}_{b,t} - (\pi_{b,t} + \alpha^{sc}_{b,t})p^{nc}_{b,t}\right] \label{opt:esr_profit_max_vl_obj}\\
& {\text{s.t.}} && \eta^c_b\sum_{t'=1}^t \left[ \sum_{v\in\mathcal{V}^{\textrm{out}}_{b,t'}} \delta_v - \sum_{v\in\mathcal{V}^{\textrm{in}}_{b,t'}} \delta_v \right] \geq \Delta\underline{s}_{b,t}+ \frac{1}{\eta^d_b} \sum_{t'=1}^t p^{nd}_{b,t'} \label{opt:esr_profit_max_vl_lb} \\
&&& \frac{\eta^c_b}{\eta^d_b}\sum_{t'=1}^t \left[\sum_{v\in\mathcal{V}^{\textrm{out}}_{b,t'}} \delta_v - \sum_{v\in\mathcal{V}^{\textrm{in}}_{b,t'}} \eta_b\delta_v \right] \leq \Delta\bar{s}_b - \eta^c_b \sum_{t'=1}^t p^{nc}_{b,t'} \label{opt:esr_profit_max_vl_ub} \\
&&& p^{nc}_{b,t} + p^{nd}_{b,t} + \sum_{v\in\mathcal{V}^{\textrm{out}}_{b,t}} \delta_v + \sum_{v\in\mathcal{V}^{\textrm{in}}_{b,t}}\eta_b \delta_v  \leq \bar{p}_b, \quad t\in\mathcal{T} \label{opt:esr_profit_max_vl_power_cap} \\
&&& \delta_b, p^{nc}_b, p^{nd}_b \geq 0
\end{align}
\end{subequations}

The profit maximization problem for ESRs \eqref{opt:esr_profit_max_vl} clearly reveals how the market clearing formulation with virtual links \eqref{opt:vl} remunerates flexibility of ESR systems. The objective function \eqref{opt:esr_profit_max_vl_obj} shows that each ESR maximizes the total profits from virtual links (shifting) and net-charging/discharging. On the first term, each virtual link is remunerated by a "discounted" price difference $\eta_{b} \pi_v^d - \pi^c_v$ across it, where the price at the discharge time is multiplied by the round-trip efficiency of the ESR. This results from the fact that each ESR has less electricity to sell than it buys when the efficiency is strictly less than 1, if the originally stored electricity is not considered. 
\\

In order to make a single virtual link $v$ profitable (thus activating the virtual link), the price difference needs to be as large as follows:
\begin{equation}
\label{eq:nonneg_profit_vl}
\eta_{b} \pi_v^d - \pi^c_v - \alpha^\delta_v \geq 0 \Rightarrow \pi^d_v - \pi^c_v \geq \frac{1-\eta_b}{\eta_b} \pi^c_v + \frac{1}{\eta_b} \alpha^\delta_v
\end{equation}
The bound reveals some interesting properties of the market design. First, the bid price plays a key role for defining the minimum price difference; a higher bid price means larger price difference is needed to activate the virtual link. Second, we observe that a higher round-trip efficiency $\eta_b$ shrinks the minimum price difference if the charging price is non-negative, but expands the difference if the charging price is very negative. 
\\

To put it more formally, let $\hat{\pi}_v := \frac{1-\eta_b}{\eta_b} \pi^c_v + \frac{1}{\eta_b} \alpha^\delta_v$ be the minimum price difference. We can compute the derivative with respect to efficiency as follows:
\begin{equation}
    \frac{\partial \hat{\pi}_v}{\partial \eta_b} = -\frac{1}{\eta_b^2}(\pi^c_v + \alpha^\delta_v)
\end{equation}
Thus, when $\pi^c_v > -\alpha^\delta_v$, more efficient ESR systems provide virtual links that can be activated with smaller price difference, while when $\pi^c_v < -\alpha^\delta_v$, less efficient ESR systems tend to provide virtual links that can be activated with smaller price difference. This can also be interpreted as evidence of how less efficient ESRs might preferred in the event of negative prices, as they can digest excess energy more easily. 
\\

In the end, we observe that the value of price $\pi^c_v$ also affects the minimum price difference $\hat{\pi}_v$. When the ESR is not ideal ($\eta_b < 1$), $\hat{\pi}_v$ monotonically increases with $\pi^c_v$. This shows that, {\em in order to activate the virtual link, the price difference has to be much larger if the prices generally reside in a high region.} On the other hand, when prices are in the negative region, even a negative price difference could be enough to incentivize the virtual link. 
\\

On the second term of the objective function \eqref{opt:esr_profit_max_vl_obj}, net-charging/discharging is remunerated via the price at the time of the operation, same as remuneration for charging/discharging shown in \eqref{opt:robust_esr_profit_max}. At each time interval $t\in\mathcal{T}$, net-discharging is incentivized when $\pi_{b,t} \geq \alpha^{sd}_{b,t} > 0$, and net-charging is incentivized when $\pi_{b,t} \leq -\alpha^{sc}_{b,t} < 0$. What makes the difference between net-charging/discharging in this virtual link formulation and charging/discharging in \eqref{opt:robust_esr_profit_max} is that net-charging/discharging variables ($p^{nc}_{b,t}$ and $p^{nd}_{b,t}$) only account for the amount of power charged/discharged that impact the net change of SOC. On the other hand, charging/discharging variables ($p^c_{b,t}$ and $p^d_{b,t}$) in \eqref{opt:robust_esr_profit_max} account for total charging/discharging power at each time, including the contribution from energy transfer over time. Indeed, formulation \eqref{opt:esr_profit_max_vl} reveals different ways ESRs react to the prices. If the prices are volatile within the time frame, an ESR generates profits by transferring power over time using virtual links; if the prices are not volatile, an ESR can choose to buy/sell power if the prices are low/high, or if there is a need to satisfy the lower and upper limits for SOC. 
\\

To complement the discussion above, we now inspect the constraints in the profit maximization formulation \eqref{opt:esr_profit_max_vl}. The constraints in the profit maximization show that there are a couple of key resources that each ESR allocates to its three types of operations, so as to achieve maximum profit: power capacity and charge capacity. The notion of power capacity is more straightforward as shown in \eqref{opt:esr_profit_max_vl_power_cap}, where at each time, the sum of power for all operations must be upper-bounded by the power limit $\bar{p}_b$. The notion of notion of charge capacity, however, might be a bit complex under the virtual link formulation. Specifically, the parameters $\Delta\bar{s}_b$/$\Delta\underline{s}_b$ define the maximum amount of net charge that can be accumulated/depleted from the ESR. 
\\

The interesting observation is that net-charging/discharging operations consume these resources permanently within the time period, while virtual links only use these resources temporarily. This is reflected on the right-hand side of constraints \eqref{opt:esr_profit_max_vl_lb} and \eqref{opt:esr_profit_max_vl_ub}, where the summation terms of $p^{nd}$ and $p^{nc}$ variables have the effect of tightening the bound of SOC for future times. This implies that ESRs will be more reluctant to do net-charging/discharging at earlier times as it might take away the capacity for virtual links to generate more profits. 

\section{Case Studies}
\label{sec:cases}
In this section, we present a couple of case studies, with increasing complexity, to demonstrate the properties of the proposed market design. Code for all case studies is available at \url{https://github.com/zavalab/JuliaBox/tree/master/StorageVirtualLink}.

\subsection{3-time, single-node system}
We consider a simple hourly market with a single spatial node (no transmission network) over a 3-hour time period. The purpose of this small case is to demonstrate theoretical properties established before, and to highlight some empirical observations that will be relevant later in cases with higher complexity. 
\\

The single node is connected to one load, one generator, and one ESR system. The amount of load requested is $\bar{d} = [25, 100, 25]$ MWh over the time period, with bid prices $\alpha^d = [30, 60, 40]$ \$/MWh. The generator has a capacity of $\bar{p} = [50,50,50]$ MWh and bid prices $\alpha^p = [5, 20, 10]$ \$/MWh. In order to create cases with negative prices, we enforce a ramping constraint for the generator:
\begin{equation}
    | p_{t+1} - p_t | \leq \Delta p
\end{equation}
where $\Delta p$ is the maximum ramp limit. The ESR we consider has charging efficiency $\eta^c = 0.9$, discharging efficiency $\eta^d = 0.8$, SOC bounds $\underline{s}_b = 0$ MWh, $\bar{s}_b = 100$ MWh, and $\bar{p}_b = 10$ MW.
\\

To demonstrate the properties of our market design, we consider four different scenarios with varying ramping capability of the generator and different initial SOC values. The parameter values of each scenario is tabulated in table \ref{table:param_case1}. In scenario 1, we assume the generator has ample ramping capability (25 MW) and the ESR starts with half of its charge capacity (50 MWh). In scenarios 2 and 4, we reduce the ramping capability to 15 MW and 5 MW to induce negative prices and observe how the ESR system reacts. In scenario 3, we increase the $s_0$ value so that the SOC upper bound could potentially be hit at the optimal solution.
\begin{table*}[ht!]
    \caption{Parameter table for single-node case study.}
    \label{table:param_case1}
    \begin{center}
    \begin{tabular}{c|cc}
     \hline
        Scenario & $\Delta p$ & $s_0$  \\
     \hline
        1 & 25 & 50 \\
        2 & 15 & 50 \\
        3 & 15 & 95 \\
        4 & 5 & 50 \\
     \hline
    \end{tabular}
    \end{center}
\end{table*}

We ran formulations \eqref{opt:market_esr} and \eqref{opt:vl} over the four scenarios. The optimal solutions are tabulated in table \ref{table:results_case1}. Here, $\phi$ denotes the total welfare value, $\pi$ the optimal prices, $p^c/p^d$ the charging/discharging power, and $s$ the SOC level. Note that they do not necessarily correspond to decision variables in the formulations, and might be calculated from optimal solutions; for instance, for the virtual link formulation \eqref{opt:vl} $p^c/p^d$ are calculated from optimal values of $\delta$ and $p^{nc}/p^{nd}$ via \eqref{eq:operations_mapping}.

\begin{table*}[ht!]
    \caption{Numerical results of the 3-time, single-node case study.}
    \label{table:results_case1}
    \begin{center}
    \begin{tabular}{cc|ccccc}
     \hline
          Model & Scenario & $\phi$ & $\pi$ & $p^c$ & $p^d$ & $s$  \\
     \hline
          \multirow{4}{*}{\eqref{opt:market_esr}} & 1  & 3883.72 & [5, 60,10] & [10,0,3.89] &  [0,10,0] & [59,46.5,50] \\
          & 2 & 3822.0 & [-0.1, 60, -0.1] & [10,0,10] & [0,10,0] & [59,46.5,55.5] \\
          & 3 & 3708.60 & [-35,60,10] & [8.14,0,8.33] & [1.86,10,0] & [100,87.5,95] \\
          & 4 & 3422.0 & [-24.9,60,-0.1] & [10,0,10] & [0,10,0] & [59,46.5,55.5]\\
     \hline
         \multirow{4}{*}{\eqref{opt:vl}} & 1 & 3883.72 & [5, 60,10] & [10,0,3.89] & [0,10,0] & [59,46.5,50] \\
         & 2 & 3822.0 & [-0.1, 60, -0.1] & [10,0,10] & [0,10,0] & [59,46.5,55.5] \\
         & 3 & 3633.72 & [-35,60,10] & [4.44,0,9.44] & [0,10,0] & [99,86.5,95]\\
         & 4 & 3422.0 & [-0.1,60,-24.9] & [10,0,10] & [0,10,0] & [59,46.5,55.5]\\
     \hline
    \end{tabular}
    \end{center}
\end{table*}

We observe that scenario 1 produces a series of positive prices that are commonly seen in real settings. Here, the ramping constraints are not active, and both market formulations produce the same optimal solution that satisfies the complementarity constraints for ESRs. Same observations can be made in scenarios 2 and 4, where the optimal solutions of the two models agree and both satisfy the complementarity constraints. As the generator becomes more inflexible with less ramping capability, more negative prices occur, leading to higher price differences that the ESR can exploit to make money. As a result, the ESR fully uses its power capacity to do price arbitrage and to buy extra electricity. 
\\

In scenario 3 we start to observe differences in the primal optimal solutions between the market formulations. Model \eqref{opt:market_esr} delivers an optimal solution that does not satisfy the complementarity constraint at time $t = 1$, where the price become quite negative (-$35$ \$/MWh). The reason is that under market formulation \eqref{opt:market_esr}, without the complementarity constraint, the ESR will take advantage of the negative price by charging and discharging simultaneously to consume extra electricity (which is not physically realizable). On the other hand, the robust bound developed for model \eqref{opt:market_esr} rules out this possibility, and therefore the optimal solution satisfies the complementarity constraint. The robust bound is also reflected at the value of $s$ at $t=1$, which is 100 for \eqref{opt:market_esr} but 99 for \eqref{opt:vl}, showing that the robust bound on SOC is active in this case (it is preventing a solution where $s = 100$ is reached). 
\\

A simple comparison with the other scenarios show that the robust bound will only be active when the ESR operates near the SOC upper bound. As a result of the robust bound being active, the total welfare value becomes smaller. These results verify the theoretical limitations of formulation \eqref{opt:market_esr} identified in Theorem \ref{thm:complementarity_condition}, where large and negative prices are likely to lead to infeasible operations for ESRs. Note that scenario 4 shows how Theorem \ref{thm:complementarity_condition} provides a sufficient but not necessary condition for guaranteeing feasible operations, as negative prices that breaks the bound in \eqref{eq:complementarity_condition} may still co-exist with feasible operations in the primal space. 

\subsection{30-node system with one ESR}

\begin{figure*}[h!]
     \centering
     \begin{subfigure}[b]{0.49\textwidth}
         \centering
         \includegraphics[width=\textwidth]{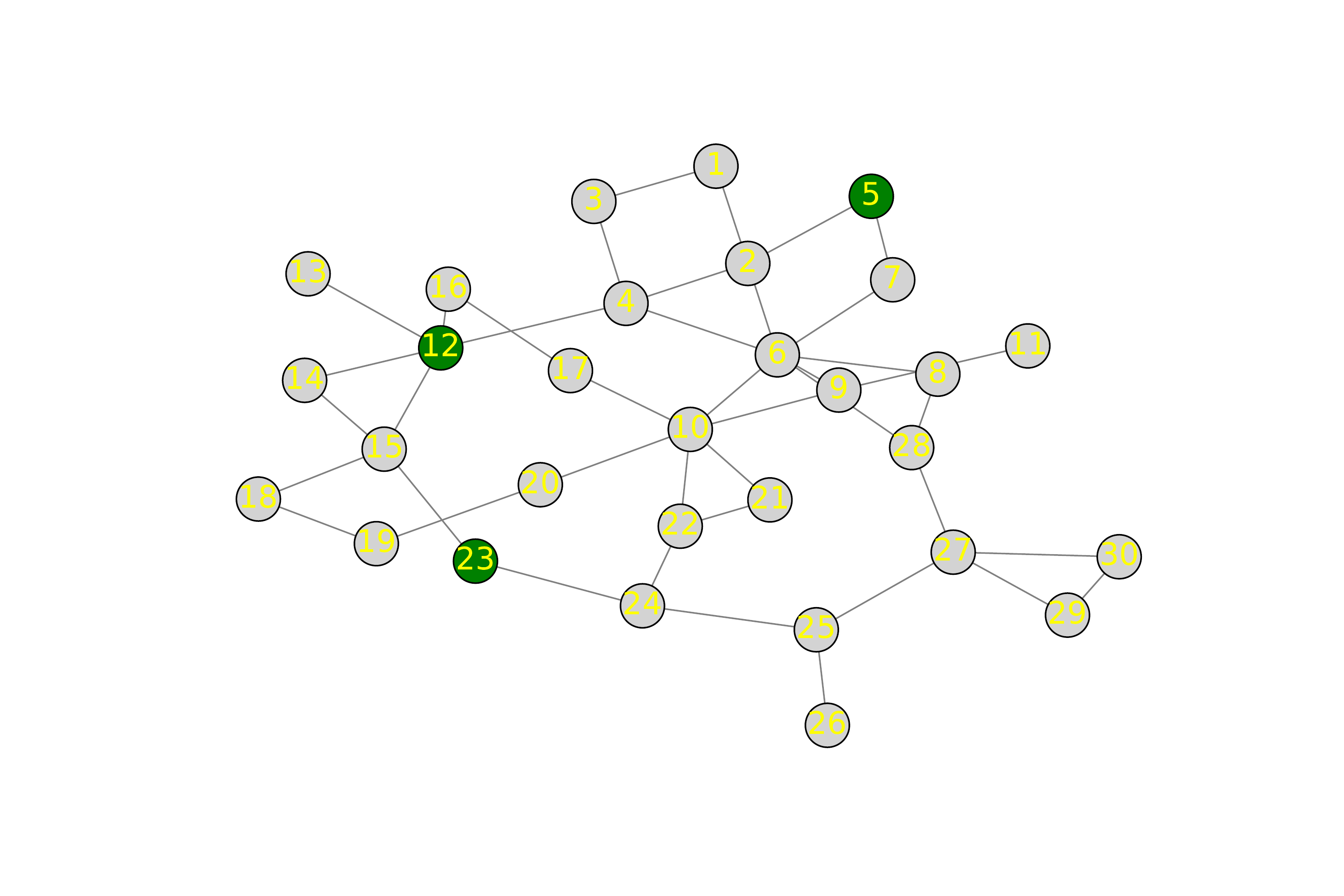}
         \caption{}
         \label{fig:ieee30_network}
     \end{subfigure}
     \hfill
     \begin{subfigure}[b]{0.49\textwidth}
         \centering
         \includegraphics[width=\textwidth]{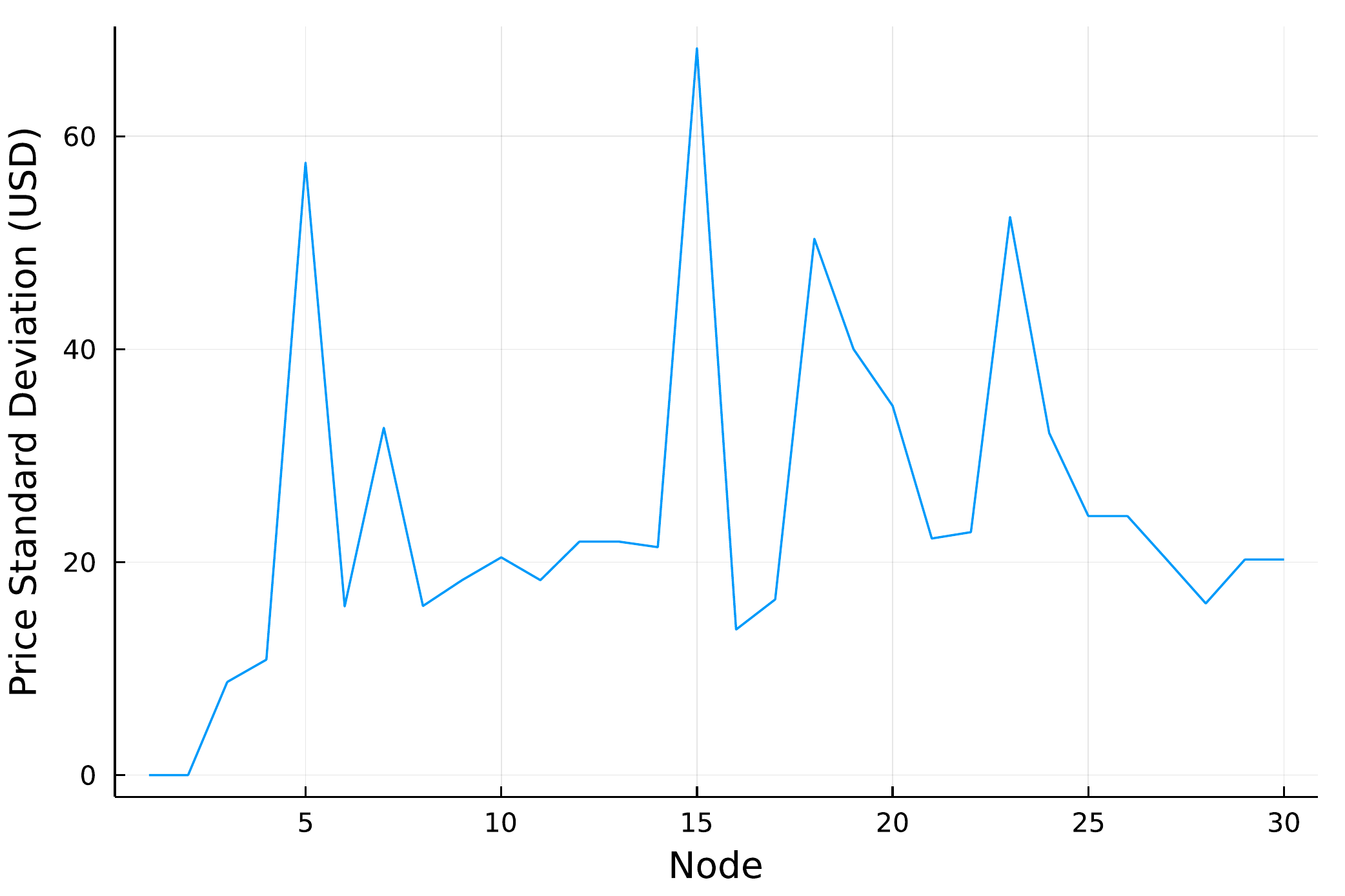}
         \caption{}
         \label{fig:base_std}
     \end{subfigure}
    \caption{\small (a) Network topology of IEEE 30-node system (each green node has an installed ESR) and (b) nodal price standard deviation under base conditions.}
    \label{fig:data}
\end{figure*}

We now consider the IEEE 30-node system with Active Power Increase (API) conditions from PGLib-OPF library \cite{babaeinejadsarookolaee2019power}. The network topology is shown in Figure \ref{fig:ieee30_network}. We ran the market over this system for a period of $T = 24$ hours. The original case file contains physical parameters for transmission lines and generators, as well as the cost function of generators. The bid cost of each generator is selected as the linear term coefficient of the cost function. The file also specifies a load level at a fixed time. To run the market over multiple hours, for each load we sample a load level for each time from a uniform distribution between 0.75 and 1.25 times of the specified load level. Each load $j$ is assigned a constant bid cost of $\alpha^d_{j,t} = 200$ \$/MWh, which resembles the value-of-loss-load (VOLL) penalty for load curtailment (note that in real life the VOLL value could be significantly higher than the 200 \$/MWh). 
\\

In this market, we consider three identical ESR systems distributed at nodes 5, 15, and 24 (green nodes in Figure \ref{fig:ieee30_network}). Each ESR has charging efficiency $\eta^c = 0.95$ and discharging efficiency $\eta^d = 0.85$. To consider the effect of varying energy and power capacity, we define an integer multiplier $K$ for the other parameters ($K=0$ are base conditions). Thus, each ESR has state of charge bounds $\underline{s}_b = 0$ MWh, $\bar{s}_b = 4K$ MWh, initial state of charge $s_0 = 2K$ MWh, and power capacity $\bar{p}_b = 1K$ MW. The multiplier $K$ takes values $\{0, 1, 5, 10, 15, 20, 25, 50\}$. The baseline scenario ($K = 0$) corresponds to the case with no ESR installed. 
\\

Figure \ref{fig:base_std} shows the temporal standard deviation of nodal prices at each node under base conditions. We observe that nodes 5 an 15 have the highest temporal price volatility over all nodes, while node 24 only exhibits medium level of volatility. In addition, we also run the case where each ESR participates individually with 3 times capacity.

\begin{figure}[h!]
    \centering
    \includegraphics[width=0.75\textwidth]{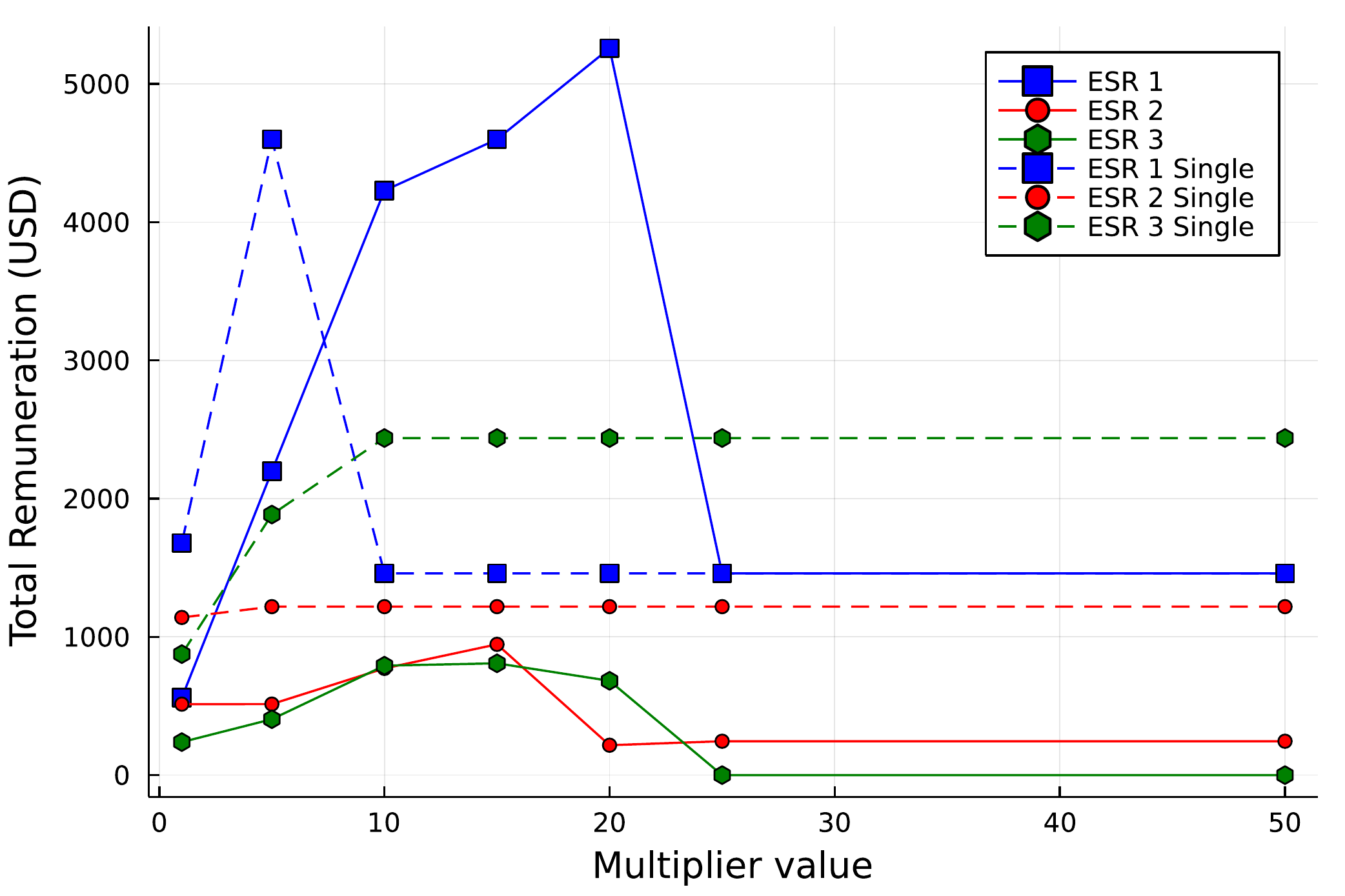}
    \caption{\small Total remuneration of each ESR for different multiplier values. Solid lines are obtained from the solution where all ESRs simultaneously participate in the market, and dashed lines ("Single" in legend) are obtained from solutions where each ESR solely participates in the market with 3 times higher capacity. }
    \label{fig:profits}
\end{figure}

Figure \ref{fig:profits} shows the total remuneration for each ESR for different multiplier values $K$. First, we observe that the choice of location has a significant effect on the economic benefits of ESR. We observe that the economic benefits of offering flexibility varies by location. Moreover, an interesting observation is that ESR 3 gains higher remuneration than ESR 2 for all cases, despite the fact that the node of ESR 2 has the highest temporal price volatility while the node of ESR 3 has only medium temporal price volatility. This implies that placing an ESR at a node where prices are volatile at the base case does not necessarily lead to high economic benefits. In fact, ESRs 1 and 2 are located at nodes with similar price volatility, but the economic potential of ESR 1 is consistently higher than that of ESR 2, in both individual and simultaneous participation cases. This gives rise to the interesting question of optimal placement of flexible technologies from an investor perspective (how to maximize economic returns). 
\\

\begin{figure}[h!]
    \centering
    \includegraphics[width=0.75\textwidth]{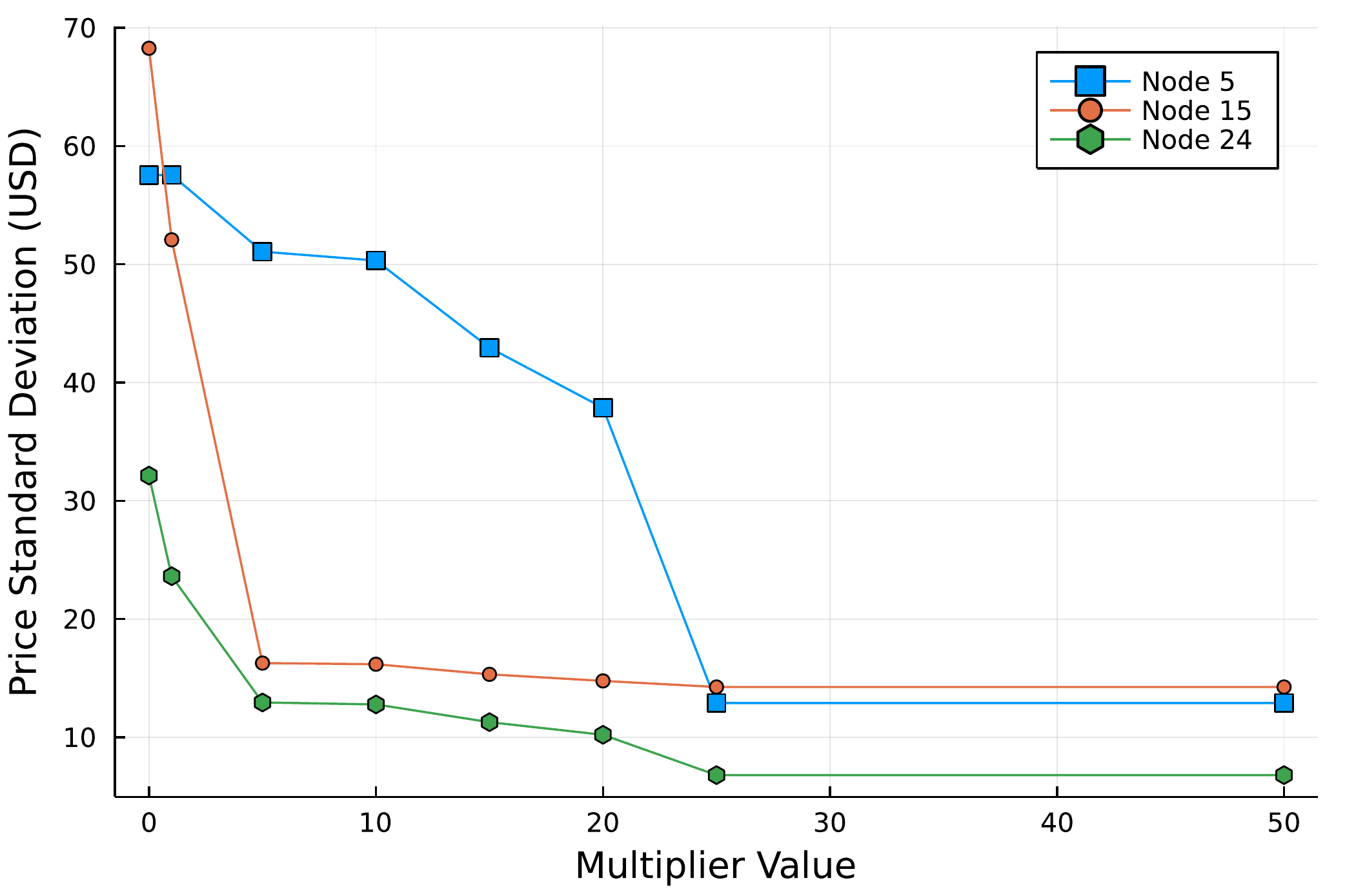}
    \caption{\small Temporal standard deviation of nodal prices at nodes 5, 15, and 24 for different values of multiplier $K$. All ESRs participate in the market simultaneously. }
    \label{fig:std_over_K}
\end{figure}

In Figure \ref{fig:profits} we observe that bidding a high capacity value does not necessarily mean higher remuneration. In most cases, the total remuneration tends to increase at smaller multiplier values and then decrease at larger multiplier values. To shed more light on this issue, Figure \ref{fig:std_over_K} shows the temporal price volatility at the three buses equipped with an ESR over different multiplier values. We observe that each node has a breaking point where price volatility sharply drops, and the break points for nodes 15 and 24 come at much smaller multiplier value compared to that of node 5. As a result, high multiplier values wipe out the economic incentives for price arbitrage that ESR system can benefit from. This explains why ESR 1 receives much higher remuneration in the simultaneous participation case, as shown in Figure \ref{fig:profits}. 
\\

All these observations demonstrate the trade-off that ESR stakeholders face when bidding in such markets: {\em bidding a small amount of flexibility might miss the chance to sell more flexibility (and get paid more), while bidding too much flexibility reduces the economic incentives and hurts the received remuneration. Therefore, ESRs need to be strategic in how much flexibility to offer.} 

\begin{figure}[h!]
    \centering
    \includegraphics[width=0.75\textwidth]{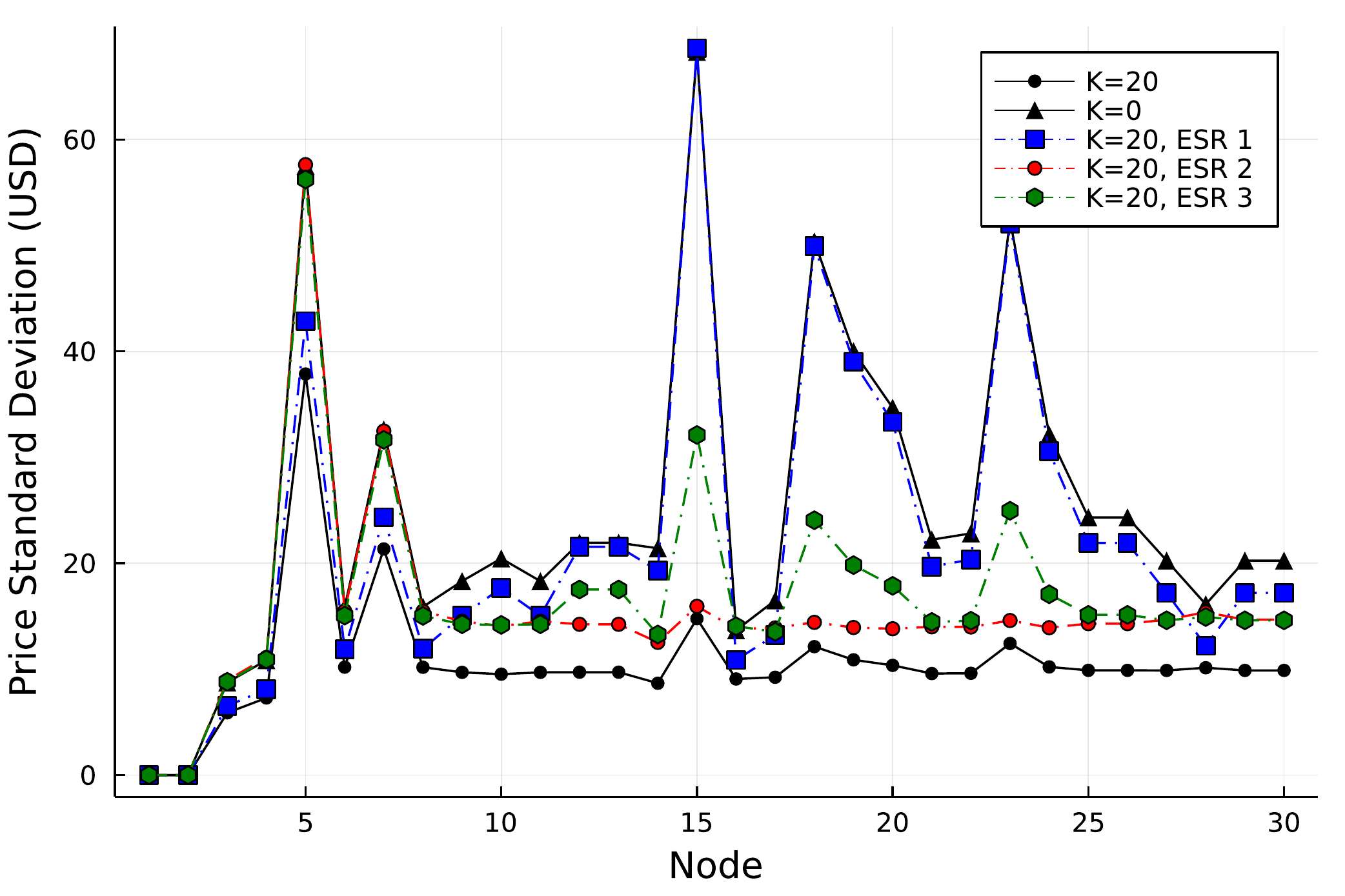}
    \caption{\small Temporal standard deviation of nodal prices for $K = 0$, $K = 20$, and $K = 20$ with individual ESR participation. The solid line with $K = 20$ denotes the case where all ESRs participate simultaneously, and the dashed lines denote the case where only one ESR participates (participating ESR is denoted in legend). }
    \label{fig:k20_std}
\end{figure}

Figure \ref{fig:k20_std} shows the temporal price volatility at different nodes for the case of $K = 0$, $K = 20$, and $K = 20$ with individual ESR participation. We observe that $K = 20$ case leads to the lowest price standard deviation, while $K = 0$ case leads to the highest price standard deviation. Cases of $K = 20$ with individual ESR participation exhibit volatility between $K = 0$ and $K = 20$. Spatial price volatility also follows the same trend. This demonstrates that from the ISO's perspective, it is more beneficial to have a larger number of  smaller ESRs distributed at different locations than to have a few large (centralized) ESRs at a single location, if the goal is to reduce system-level price volatility. Interestingly, our market does not make any assumption about the ownership of ESRs, meaning that the market framework gives the same solution regardless of the ownership of ESRs. 
\\

If each ESR belongs to a different entity, we can observe that multiple small ESRs will compete with each other to offer more flexibility,  leading to a higher reduction in price volatility with cheaper cost (less remuneration for ESRs). This can be verified partly in Figure \ref{fig:profits}, where at some multiplier values (e.g. $K = 5$), the remuneration for all ESRs combined in the simultaneous participation case is even lower than the remuneration for a single large ESR in the individual participation case. On the other hand, if all the ESRs belong to the same entity, the results imply that ESR investors may be incentivized to reduce distribution of ESR units across different locations, in order to avoid possible reduction in remuneration.

\section{Conclusions}
\label{sec:conclusion}
Recent trends of renewable energy absorption calls for much higher needs for flexibility resources. New FERC orders and regulations call for ISOs to implement new markets to allow ESR participation, but research work in this area is lacking. In this paper, we propose a new energy market clearing framework that models ESR flexibility using virtual links. We discuss the application of a robust bound for ESR operations that helps satisfy complementarity conditions without using mixed-integer formulations. We prove a theoretical lower bound for prices where the robust bound might be needed. Our market framework decomposes the operations of ESRs to reveal the incentives for ESR operations. By applying the concept of virtual links and Lagrangian dual analysis, we show that load shifting is remunerated via temporal price volatility, and efficiency of ESR plays a key role in the economic benefits of the load-shifting flexibility. We also demonstrate that the SOC capacity and power capacity are the real resources that determine how much load-shifting flexibility an ESR can provide. 

The case studies give rise to several interesting directions for future work. From an investor perspective,  the capacity sizing and optimal placement of ESR resources can be done via a bilevel programming formulation that embeds our market clearing framework. This is feasible as our market clearing framework avoids mixed-integer formulation from complementarity. From the ESR operators' perspective, it is worth exploring the optimal bidding strategy in detail. In this work we assume truthful bidding, i.e. the bid price and capacity perfectly reflect the operating cost and capacity of the ESR system. However, the case studies identify possible incentives for ESRs to bid less capacity than their actual capacity. In addition, bid price gives a lower bound on the profit of virtual links as found in \cite{zhang2021electricity}, which provides another motivation for the submission of higher bid prices. From the ISO's perspective, it is worth exploring how to extend our market framework for multi-scale markets. The current market is formulated on a day-ahead basis, and thus virtual links only incentivize load shifting in short period of time. The incentives for long-term load shifting will have to be provided in another layer of market as real-life electricity markets work on multiple time scales. One possible approach is to design another market at a higher scale (e.g., seasonal or monthly scale) that determines the operating SOC level (which is a parameter in the current market framework). The change of SOC levels at different seasons/months can also be capture via the concept of virtual links.


\section*{Acknowledgments}
We acknowledge support from the U.S. National Science Foundation under award 1832208.


\bibliography{refs}


\appendix
\appendixpage
\section{Proofs}

\begin{proof}[Proof of Theorem \ref{thm:complementarity_condition}]
We inspect the profit maximization problem \eqref{opt:esr_profit_max} for each ESR $b \in \mathcal{B}$. Given $\pi$, let $(p^c_b, p^d_b)$ be a feasible solution of \eqref{opt:esr_profit_max}. For convenience, let $\psi(p^c_b, p^d_b)$ denote the objective function \eqref{opt:esr_profit_max_obj}. Define $\epsilon_t := \min\{p^c_{b,t}, \frac{1}{\eta_b} p^d_{b,t}\} \geq 0$ for every $t \in \mathcal{T}$, and $\epsilon := \{\epsilon_t\}_{t\in\mathcal{T}}$. First, we observe that $(p^c_b - \epsilon, p^d_b - \eta_b \epsilon)$ is also a feasible solution to \eqref{opt:esr_profit_max} because
\begin{equation*}
    \eta^c_b \sum_{t' = 1}^t (p^c_{b,t'}-\epsilon_{t'}) - \frac{1}{\eta^d_b} \sum_{t' = 1}^t (p^d_{b,t'}-\eta_b\epsilon_{t'}) = \eta^c_b \sum_{t' = 1}^t p^c_{b,t'} - \frac{1}{\eta^d_b} \sum_{t' = 1}^t p^d_{b,t'} + \sum_{t'=1}^t\eta^c_b(\epsilon_{t'} - \epsilon_{t'}) = \eta^c_b \sum_{t' = 1}^t p^c_{b,t'} - \frac{1}{\eta^d_b} \sum_{t' = 1}^t p^d_{b,t'}
\end{equation*}
for every $t \in \mathcal{T}$. Next we observe that $(p^c_b - \epsilon, p^d_b - \eta_b \epsilon)$ satisfies the complementarity condition. This is true because by definition of $\epsilon$, at every time $t$ either $p^c_{b,t} = 0$ or $p^d_{b,t} = 0$. In the end, we show that $(p^c_b - \epsilon, p^d_b - \eta_b \epsilon)$ produces a better objective value than $(p^c_b, p^d_b)$ under the condition of \eqref{eq:complementarity_condition}, meaning that the optimal solution must satisfy the complementarity condition. 
\begin{equation*}
\begin{aligned}
    & \psi(p^c_b - \epsilon, p^d_b - \eta_b \epsilon) - \psi(p^c_b, p^d_b) \\
    = & \sum_{t \in \mathcal{T}} (\pi_{b,t} - \alpha^{sd}_{b,t})(p^d_{b,t}-\eta_b \epsilon_t) - (\pi_{b,t} + \alpha^{sc}_{b,t})(p^c_{b,t}-\epsilon_t) - (\pi_{b,t} - \alpha^{sd}_{b,t})p^d_{b,t} + (\pi_{b,t} + \alpha^{sc}_{b,t})p^c_{b,t} \\
    = & \sum_{t \in \mathcal{T}} \eta_b \alpha^{sd}_{b,t} \epsilon_t + \alpha^{sc}_{b,t} \epsilon_t + \epsilon_t (1-\eta_b)\pi_{b,t} \\
    \geq & \sum_{t \in \mathcal{T}} \epsilon_t \left(\eta_b \alpha^{sd}_{b,t} + \alpha^{sc}_{b,t} + (1-\eta_b)\frac{-1}{1-\eta_b}(\eta_b\alpha^{sd}_{b,t} + \alpha^{sc}_{b,t})\right) \\
    = & \, 0 
\end{aligned}
\end{equation*}
\end{proof}

\begin{proof}[Proof of Theorem \ref{thm:robust_complementarity}]
We inspect the profit maximization problem \eqref{opt:robust_esr_profit_max} for each ESR $b \in \mathcal{B}$. Given $\pi$, let $(p^c_b, p^d_b)$ be a feasible solution of \eqref{opt:robust_esr_profit_max}. For convenience, let $\psi(p^c_b, p^d_b)$ denote the objective function \eqref{opt:robust_esr_profit_max_obj}. Define $\epsilon_t := \min\{p^c_{b,t}, p^d_{b,t}\} \geq 0$ for every $t \in \mathcal{T}$, and $\epsilon := \{\epsilon_t\}_{t\in\mathcal{T}}$. First, we observe that $(p^c_b - \epsilon, p^d_b - \eta_b \epsilon)$ is also a feasible solution to \eqref{opt:esr_profit_max} because
\begin{equation*}
    \eta^c_b \sum_{t' = 1}^t (p^c_{b,t'}-\epsilon_{t'}) - \frac{1}{\eta^d_b} \sum_{t' = 1}^t (p^d_{b,t'}-\epsilon_{t'}) = \eta^c_b \sum_{t' = 1}^t p^c_{b,t'} - \frac{1}{\eta^d_b} \sum_{t' = 1}^t p^d_{b,t'} + \sum_{t'=1}^t(\frac{1}{\eta^d_b}-\eta^c_b)\epsilon_{t'} \geq \eta^c_b \sum_{t' = 1}^t p^c_{b,t'} - \frac{1}{\eta^d_b} \sum_{t' = 1}^t p^d_{b,t'}
\end{equation*}
for every $t \in \mathcal{T}$.

Next we observe that $(p^c_b - \epsilon, p^d_b - \epsilon)$ satisfies the complementarity condition. This is true because by definition of $\epsilon$, at every time $t$ either $p^c_{b,t} = 0$ or $p^d_{b,t} = 0$. In the end we show that $(p^c_b - \epsilon, p^d_b - \epsilon)$ produces a better objective value than $(p^c_b, p^d_b)$ under the condition of \eqref{eq:complementarity_condition}, meaning that the optimal solution must satisfy the complementarity condition. 
\begin{equation*}
\begin{aligned}
    & \psi(p^c_b - \epsilon, p^d_b - \epsilon) - \psi(p^c_b, p^d_b) \\
    = & \sum_{t \in \mathcal{T}} (\pi_{b,t} - \alpha^{sd}_{b,t})(p^d_{b,t}-\epsilon_t) - (\pi_{b,t} + \alpha^{sc}_{b,t})(p^c_{b,t}-\epsilon_t) - (\pi_{b,t} - \alpha^{sd}_{b,t})p^d_{b,t} + (\pi_{b,t} + \alpha^{sc}_{b,t})p^c_{b,t} \\
    = & \sum_{t \in \mathcal{T}} (\alpha^{sd}_{b,t} + \alpha^{sc}_{b,t})\, \epsilon_t \\
    \geq & \, 0 
\end{aligned}
\end{equation*}
\end{proof}

\begin{proof}[Proof of Theorem \ref{thm:vl_equivalence}]
To establish equivalence, we show the solution of one formulation can be obtained from some solution of the other formulation with the same objective value. For that we inspect the linear system \eqref{eq:operations_mapping}. For convenience, we write the system as $y = Ax$, where $y := [p^c_b, p^d_b]$ captures the solution of \eqref{opt:robust_market_esr}, and $x := [\delta, p^{nc}_b, p^{nd}_b]$ captures the solution of \eqref{opt:vl}. Trivially, one can obtain $y$ from an optimal $x$ following \eqref{eq:operations_mapping}. Thus, for the rest of the proof we focus on how to obtain a solution $x$ given $y$. Let $\mathcal{X}_y$ be the set of solutions to the system \eqref{eq:operations_mapping} given $y$. Note that matrix $A$ is full row rank as the identity matrix $\mathbf{I}_{2T, 2T}$ is a submatrix of $A$. This implies that given an arbitrary $y$, the solution set $\mathcal{X}_y$ is non-empty (one trivial solution would be to set $p^{nc}_b = p^c_b$, $p^{nd}_b = p^d_b$).
\\

We now show that there exists $x \in \mathcal{X}_y$ that is feasible to constraints \eqref{opt:vl_soc_lb}-\eqref{opt:vl_nonneg} given $y$ from a feasible solution of \eqref{opt:robust_market_esr}. \eqref{opt:vl_power_cap} holds naturally for every $x \in \mathcal{X}_y$ as implied by the fact that $y$ satisfies constraint \eqref{opt:robust_market_esr_power_cap}. To show how $x$ can be guaranteed to satisfy other constraints, we write down a recipe of how to find the solution. We first compute the net change of SOC of the ESR given $y$:
\begin{equation}
    \Delta \textrm{SOC} := \eta^c_b \sum_{t=1}^t p^c_{b,t} - \frac{1}{\eta^d_b} \sum_{t=1}^t p^d_{b,t}
\end{equation}
If $\Delta \textrm{SOC} = 0$ then the ESR is purely shifting energy within the time window, meaning that we can set $p^{nc}_{b,t} = 0$ and $p^{nd}_{b,t} = 0$. If $\Delta \textrm{SOC} > 0$/$\Delta \textrm{SOC} < 0$ then solution $y$ induces a net-charging/discharging behavior, so some of the $p^{nc}_{b,t}$/$p^{nd}_{b,t}$ are set with strictly positive values. The values are assigned such that $p^{nd}_{b,t} = 0$ and $\eta^c_b \sum_{t\in\mathcal{T}} p^{nc}_{b,t} = \Delta \mathrm{SOC}$ if $\Delta \textrm{SOC} > 0$, or $p^{nc}_{b,t} = 0$ and $\frac{1}{\eta^d_b} \sum_{t\in\mathcal{T}} p^{nd}_{b,t} = -\Delta \mathrm{SOC}$ if $\Delta \textrm{SOC} < 0$. Once $p^{nc}_{b,t}, p^{nd}_{b,t}$ are determined, we are left with a time graph, where each time node is either a source node defined by $p^c_{b,t} - p^{nc}_{b,t}$ or a sink node defined by $p^d_{b,t} - p^{nd}_{b,t}$. By construction we have 
\begin{equation}
    \eta_b \cdot \left(\sum_{t\in\mathcal{T}}p^c_{b,t} - p^{nc}_{b,t}\right) = \sum_{t\in\mathcal{T}} p^d_{b,t} - p^{nd}_{b,t}
\end{equation}
so the charge into and out of the ESR is balanced. Then, a feasible $\delta$ can be easily found by repetitively pairing source and sink nodes until all nodes have zero net energy. With this procedure, one can easily verify that the resulting solution $x$ satisfies constraints \eqref{opt:vl_soc_lb}-\eqref{opt:vl_nonneg} by the feasibility of $y$ in \eqref{opt:robust_market_esr}.

In the end, we show that the new solution gives the same objective value. Let $(x,y)$ be a pair of solutions to $y = Ax$. We have that:
\begin{equation*}
\begin{aligned}
    & \sum_{t\in\mathcal{T}}\alpha^{sc}_{b,t} p^c_{b,t} + \alpha^{sd}_{b,t} p^d_{b,t} \\
    = \, & \sum_{t\in\mathcal{T}} \alpha^{sc}_{b,t}\left(\sum_{v\in\mathcal{V}^{\textrm{out}}_{b,t}}\delta_v + p_{b,t}^{nc}\right) + \alpha^{sd}_{b,t}\left(\sum_{v\in\mathcal{V}^{\textrm{in}}_{b,t}}\eta_{b(v)} \delta_v + p_{b,t}^{nd}\right) \\
    = \, & \sum_{t\in\mathcal{T}} \left(\alpha^{sc}_{b,t} p_{b,t}^{nc} + \alpha^{sd}_{b,t}p_{b,t}^{nd}\right) + \sum_{v\in\mathcal{V}} (\alpha^{sc}_{b,t} + \eta_{b(v)}\alpha^{sd}_{b,t})\delta_v \\
    = \, & \sum_{t\in\mathcal{T}} \left(\alpha^{sc}_{b,t} p_{b,t}^{nc} + \alpha^{sd}_{b,t}p_{b,t}^{nd}\right) + \sum_{v\in\mathcal{V}} \alpha^\delta_v \delta_v
\end{aligned}
\end{equation*}
This concludes the proof.
\end{proof}

\end{document}